\documentclass[12pt]{iopart}
\usepackage{graphicx}
\usepackage{dcolumn}
\usepackage{bm}
\usepackage{tabularx}
\usepackage{ulem}
\newcolumntype{M}{>{\centering\arraybackslash}m{1.85cm}}
\usepackage[export]{adjustbox}
\usepackage{float}
\usepackage{braket}
\usepackage{makecell}
\usepackage{lipsum}
\usepackage{longtable}
\graphicspath{{figure/}}
\usepackage{xcolor}   

 \usepackage{hyperref}

\begin{document}

\title{Systematic shell-model study of structure and isomeric states in $^{204-213}$Bi isotopes}

\author{Sakshi Shukla, Praveen C. Srivastava,\footnote{Corresponding author:  praveen.srivastava@ph.iitr.ac.in} Deepak Patel }
\address{Department of Physics, Indian Institute of Technology Roorkee, Roorkee
	247 667, India}

\vspace{10pt}

\begin{abstract}
In this work, we have performed systematic  shell-model calculations for Bi isotopes with $A=$ 204-213  using KHH7B and KHM3Y effective interactions. We have reported yrast and non-yrast shell-model states corresponding to the available experimental data. From the comparison with the experimental data, we could assign spin and parity of several unconfirmed states. We have also calculated electromagnetic properties and compared them with the available experimental data and predicted where experimental data are not available. This study also includes a detailed discussion of multiple isomeric states based on computed shell-model configurations and their respective half-lives.
\end{abstract}

%
%
%
%
%

\section{Introduction}

\begin{figure}
	\begin{center}
		\includegraphics[width=10.65cm]{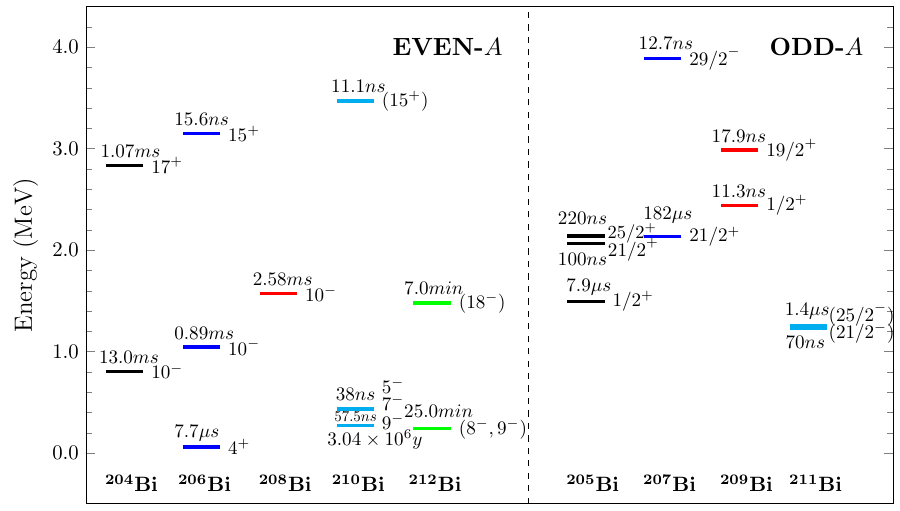}
	\end{center}
	\caption{\label{isomer} Experimentally observed isomeric states and their half-lives for $^{204-212}$Bi.}
\end{figure}

There are numerous experimental data available near doubly closed core $^{208}_{82}$Pb$_{126}$ nucleus \cite{broda, Barzakh, Wahid, podolyak, Das}. Thus, it will be very useful to test shell-model (SM) effective interactions in this region. Also, the $^{208}$Pb region offers a chance to examine and explore complex nuclear structures such as shell evolution, shape changes  and octupole collectivity \cite{Ojala, Rout, Brown, Piet}. In this mass region, quadrupole collectivity is found in different rotational bands of many nuclei \cite{Morrison, Herzan}; studying such important properties helps us to understand nuclear deformation. Moreover, the study of the behavior of nuclei near $N=126$ also gives the understanding of the astrophysical $r$-process in the formation of heavier nuclei \cite{Nieto, r-process}. Recently, our group studied the collectivity and isomerism in the Pb and Po isotopes \cite{Sakshi2, Sakshi1}. Now, our focus is to study Bi isotopes that lie in the neighborhood of neutron shell closure $N=126$. In  $^{204-213}$Bi isotopes, we can see interactions between one valence proton and several neutrons below and above the $N=126$. Among the $^{204-213}$Bi nuclei, the $^{208,210}$Bi are more intriguing for the theoretical analysis because their lower energy states likely result from the interaction between a valence proton with a neutron hole in $^{208}$Bi nucleus, and similarly in $^{210}$Bi one valence proton with one valence neutron particle just above the $N=126$ \cite{Boutachkov,sheline}. The other isotopes are also interesting to study proton-neutron excitation and coupling between them with core excitations. The nuclear shell model for Bismuth appears to be the most suitable model for defining the specific nuclear structure behavior since it only has a small number of nucleons outside the doubly magic $^{208}$Pb nucleus. There are enormous amount of data available for Bismuth isotopes that helps us to give theoretical predictions for energy levels, half-lives of different isomers, and electromagnetic properties \cite{Hopke, Cieplicka, Bieron, Chen}. Also, it is possible to examine the emergence of an island of extreme nuclear isomerism at high excitation in the Pb region \cite{garg,Wahid_PLB,kyadav}. In Fig. \ref{isomer}, the experimentally observed isomers with their half-lives for Bi isotopes are shown. In the present study, we intend to explore these isomeric states using the shell-model.

Many experiments have been performed at RIKEN, GSI/FAIR, and CERN to measure the energy levels and the electromagnetic properties in the Pb region to understand nuclear structure. The lack of understanding of their nuclear structure is mostly due to experimental difficulty in generating isotopes above the shell closure ($\sim$ $N=126$) in the Pb region. In the previous studies, Byrne \textit{et al.} \cite{byrne} investigated the high-spin states in the $^{205}$Bi nucleus by utilizing gamma-ray spectroscopic methods,  which were populated through the $^{205}$Tl($\alpha$,4n) reaction at approximately 50 MeV bombarding energies. Notably, states with spins reaching up to 39/2$^+$ and excitation energies of roughly 6 MeV were successfully identified.  Earlier, following the deep-inelastic reactions and using $\gamma$-ray coincidence spectroscopy involving $^{76}$Ge + $^{208}$Pb, new microsecond isomers with high spin values of $J^\pi$ = (31$^+$) and $J^\pi$ = (28$^-$) in $^{206}$Bi at energy levels of 10,170 keV and 9,233 keV, respectively have been reported in Ref. \cite{Cieplicka}. High spin states of $^{207}$Bi using the reaction $^{205}$Tl($\alpha$, 2n)$^{207}$Bi and $^{208}$Pb(d, 3n)$^{207}$Bi were studied utilizing $\alpha$-particle in the energy range 30-43 MeV, and deuteron with energy 25 MeV in Ref. \cite{lonnroth}. In the case of $^{212}$Bi isotope, a long-lived isomer $(18^-)$ with half-life 7.0(3) min  at 1478(30) keV  was observed \cite{Chen}.   As we mentioned above $^{208}$Bi exhibits a single neutron vacancy and an additional proton  relative to the doubly magic $^{208}$Pb,  making it a prime example of the particle-hole interaction. Alford \textit{et al.} \cite{Alford} conducted measurements involving the $^{207}$Pb($^3$He, d)$^{208}$Bi and $^{207}$($\alpha$, t)$^{208}$Bi reactions, leading to the excitation of levels featuring a $p_{1/2}$ neutron vacancy. They also conducted the experiment involving $^{209}$Bi(d, t)$^{208}$Bi and $^{209}$Bi($^3$He, $\alpha$) reactions, which shows  proton presence in the $h_{9/2}$ orbital. Martin \cite{martin} evaluated the experimental data for this isotope previously. Recently, using in-source laser spectroscopy, the isotope shift with respect to $^{209}$Bi and the hyperfine splitting for $^{211,213}$Bi have been detected at the atomic transition of 306.77 nm \cite{Barzakh}. It has been noted that there is a noticeable shell effect in the radii and magnetic moments of Bi isotopes at $N=126$.  The isotopic trend for the magnetic moment of the 9/2$^-_{g.s.}$ in odd-$A$ Bi isotopes can be ascribed to the change in the first-order core-polarization correction.

Extensive theoretical investigations within this mass range have been conducted \cite{Mcgrory, Coraggio, Caurier, koji, Teruya, Yanase, Naidja, Wilson, Anil}. Several decades ago, McGrory and Kuo \cite{Mcgrory}  performed shell-model calculations to explore the nuclear structure of $^{204-206}$Pb, $^{210-212}$Pb, $^{210}$Po, $^{211}$At, and $^{212}$Rn nuclei, which are characterized by a limited number of valence nucleons around the $^{208}$Pb core. 
 In the present work we have performed systematic shell-model calculations for $^{204-213}$Bi isotopes in the extended model spaces. Recently there are numerous experimental data are available for energy spectra, electromagnetic properties, and isomeric states. To the best of our knowledge, there is no systematic theoretical study available for the Bi chain in the literature. Thus we aim to report theoretical results corresponding to these new data. Additionally we have discussed about the deformation for the ground states of the odd-$A$ Bi isotopes. Also, there are numerous experimental activities  to measure beta decay properties at CERN in this mass region including Bi isotopes \cite{cern}. Thus present nuclear structure study demonstrate that it is possible to use these wavefunctions to predict theoretical results for beta decay properties corresponding to existing and upcoming experimental data.

This paper is organized as follows: Section \ref{II} provides a  brief introduction of the effective shell-model interaction used in the present calculation. Section \ref{III} presents the outcomes of the shell-model analysis. Finally, Section \ref{IV} summarizes the findings and draws conclusions.

\section{Formalism: Model Space and Hamiltonian}\label{II}

We can express the shell-model Hamiltonian numerically using single-particle energies and the two-body matrix elements (TBMEs) as,
\begin{equation}
	H=\sum_{\alpha}\varepsilon_{\alpha}{\hat N}_{\alpha}+\frac{1}{4}\sum_{\alpha\beta\delta\gamma JT}\langle j_{\alpha}j_{\beta}|V|j_{\gamma}j_{\delta}\rangle_{JT}A^{\dag}_{JT;j_{\alpha}j_{\beta}} \times 
	A_{JT;j_{\delta}j_{\gamma}},
\end{equation}
here, $\alpha=\{nljt\}$ represents the single-particle orbitals and $\varepsilon_{\alpha}$ refers to the corresponding single-particle energies. $\hat{N}_{\alpha}=\sum_{j_z,t_z}a_{\alpha,j_z,t_z}^{\dag}a_{\alpha,j_z,t_z}$  stands for the particle number operator. The two-body matrix elements
$\langle j_{\alpha}j_{\beta}|V|j_{\gamma}j_{\delta}\rangle_{JT}$ are coupled to the spin $J$ and isospin $T$. $A_{JT}^{\dag}$, and $A_{JT}$  denote the fermion pair creation and annihilation operators, respectively.

Shell-model calculations have been performed utilizing the KHH7B and KHM3Y interactions. Model space of  KHH7B interaction \cite{popellier} have 14 orbitals, and the cross-shell TBMEs are generated from the H7B G-matrix \cite{h7b}, while the neutron-proton TBMEs are established with the application of Kuo-Herling interaction \cite{kuo} as modified in \cite{kuo_mod}. The second interaction taken here is KHM3Y; its model space consists of 24 orbitals. The M3Y interaction \cite{m3y} provides the foundation for the cross-shell TBMEs for the KHM3Y interaction, while the neutron-proton interactions are derived from the Kuo-Herling interaction \cite{kuo} as modified in Ref. \cite{kuo_mod}.

To make calculations feasible, we have applied appropriate truncation. For KHH7B model space, we have completely filled proton orbitals below Z=82 and allowed one valance proton to occupy any one of the orbitals above Z=82, namely in $0h_{9/2}$, $1f_{7/2}$, and $0i_{13/2}$. We have allowed valence neutrons to occupy $1f_{5/2}$, $2p_{3/2}$, $2p_{1/2}$, and $0i_{13/2}$ orbitals below $N=126$ for $^{204-208}$Bi nuclei. For $^{209}$Bi, $1p-1h$ excitation applied across $N=126$. While for $^{210-213}$Bi isotopes, we have allowed only those neutrons to fill into the $1g_{9/2}$, $0i_{11/2}$, and $0j_{15/2}$ orbitals which are above $N=126$. For KHM3Y model space, for valance protons, we have completely filled all the five orbitals below $Z=82$, and allowed one proton to occupy into any of the six $0h_{9/2}$, $1f_{7/2}$, $0i_{13/2}$, $1f_{5/2}$, $2p_{3/2}$, and $2p_{1/2}$ orbitals above $Z=82$. For valence neutrons, restrictions are such that only orbitals below $N=126$ ($0h_{9/2}$,  $1f_{7/2}$, $1f_{5/2}$, $2p_{3/2}$, $2p_{1/2}$, and $0i_{13/2}$) are allowed for $^{204-208}$Bi isotope.
Due to computational challenges, we have applied some truncation on the neutron orbitals for $^{204}$Bi, where all the neutron partitions belonging to the $\nu(h_{9/2}^{8-10}i_{13/2}^{0-14}f_{7/2}^{6-8}p_{3/2}^{0-4}f_{5/2}^{0-6}p_{1/2}^{0-2})$ considered. Whereas for $^{209-212}$Bi, seven neutron orbitals $1g_{9/2}$, $0i_{11/2}$, $0j_{15/2}$, $2d_{5/2}$, $3s_{1/2}$, $1g_{7/2}$, and $2d_{3/2}$ above $N=126$ are also open. In $^{209}$Bi, $1p-1h$ excitation employed for neutrons at $N=126$. For $^{210-212}$Bi isotopes, we have allowed only those valence neutrons to fill into these seven orbitals, which are above $N=126$. In the case of $^{213}$Bi isotope due to huge dimension, we have allowed four valance neutrons which are above the $N=126$ to occupy in only three neutron orbitals $1g_{9/2}$, $0i_{11/2}$, and $0j_{15/2}$. Recently, our group employed KHH7B and KHM3Y effective interactions in the following Refs. \cite{bharti,chatterjee,Das,Bharti2}. Shell-model calculations in the extended model space (with KHM3Y interaction) compared to the previous studies \cite{Yanase, T.Lonnroth} can be helpful in investigating the different spectroscopic properties of Bi isotopes. To diagonalize the shell-model Hamiltonian matrices, we have used KSHELL \cite{kshell} and NUSHELLX \cite{nushellx}. All the reported results of the KHH7B interaction are obtained utilizing the KSHELL code. Similarly, the results of the KHM3Y interaction are calculated using the NUSHELLX code.


\section{Results and Discussions\label{III}}

Here, we have discussed the energy spectra and the reduced transition probabilities for even and odd-$A$ Bi isotopes with $A=204-213$ in sections \ref{even-A}, and \ref{odd-odd}, respectively. The comparison between the experimental and shell-model predicted $B(E2)$ transitions are reported in Table \ref{be2}. The electromagnetic moments, such as magnetic and quadrupole moments, have been discussed in section \ref{EM}. The half-lives and seniority of isomeric states are reported in section \ref{isomer1}. The root mean square (rms) deviation between the experimental and calculated $B(E2)$ values, quadrupole, and magnetic moments are discussed in section \ref{rms_deviation}.

\subsection{\label{even-A} Results for even-$A$ $^{204-212}$Bi isotopes}

\begin{figure}
	\centering
	\includegraphics[width=8.50cm,height=9cm]{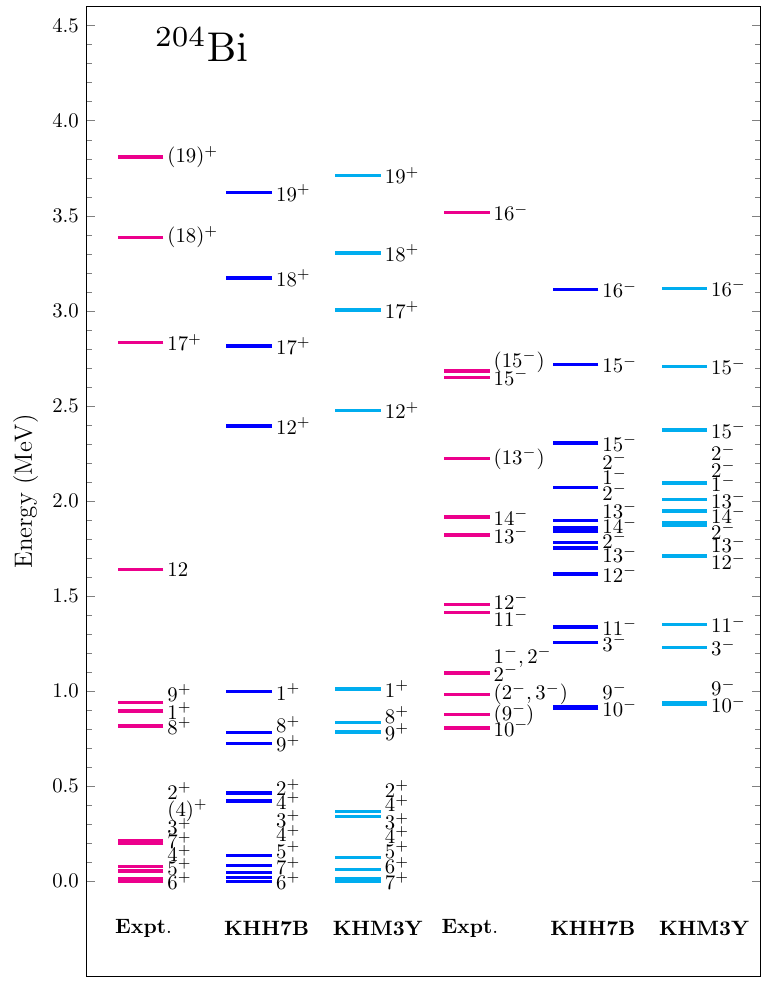}
	\caption{\label{204Bi} Comparison between the theoretical and experimental \cite{NNDC} energy levels in $^{204}$Bi isotope.}
	
\end{figure}

{\bf $^{204}$Bi}:  The comparison between the theoretical and experimental energy spectra of $^{204}$Bi using the KHH7B and KHM3Y interactions is depicted in Fig. \ref{204Bi}. We have reported only yrast and non-yrast states that correspond to the available experimental data. The yrast $2^+-8^+$ states are obtained by the same dominant configuration $\pi(h_{9/2}^1)\otimes\nu (f_{5/2}^3p_{3/2}^4i_{13/2}^{14})$. For this isotope, 6$^+$ is the ground state (g.s.) experimentally, which is reproduced theoretically by KHH7B interaction only. For positive parity states above excitation energy 2.5 MeV, KHH7B interaction shows good results for the $17^+_1$ state, whereas $18^+_1$ and $19^+_1$ states are reproduced reasonably well by KHM3Y interaction. For negative parity states, we are also able to reproduce the first yrast state 10$^-$ from both interactions, and its configuration is $\pi(h_{9/2}^1)$  $\otimes $  $\nu (f_{5/2}^4p_{3/2}^4i_{13/2}^{13})$ using KHH7B interaction with 49.12 $\%$ probability. The other states ($11^-$ to $15^-$) of the negative-parity yrast cascade are also characterized by same configuration [$\pi(h_{9/2}^1)$  $\otimes $  $\nu (f_{5/2}^4p_{3/2}^4i_{13/2}^{13})$] and experimentally connected with the magnetic dipole $(M1)$ transition \cite{Lonroth}. Although the experimental $B(M1)$ values are not observed yet, the shell model predicted values for $15^-_1\rightarrow14^-_1$, $14^-_1\rightarrow13^-_1$, $13^-_1\rightarrow12^-_1$, $12^-_1\rightarrow11^-_1$, and $11^-_1\rightarrow10^-_1$ transitions are 0.528, 0.145, 0.262, 0.110, and 0.583 $\mu_N^2$, respectively; which might be useful to compare the upcoming experimental data in the future. Our results support the experimentally tentative state $(9^-)$ at 0.876 MeV to be the 9$^-_1$ state. Experimentally, at 0.983 MeV, a tentative state (2$^-$,3$^-$) is obtained; further,  by comparing this state with theoretical results, we see that $3^-_1$ and $2^-_1$ states lie 271 keV and 772 keV above, respectively. Thus, we can predict this state belongs to 3$^-_1$. The tentative experimental spin parity state at 2.684 MeV excitation energy is $(15^-)$, which can be assigned as $15^-_2$ state by shell-model calculations. For better improvement of the high-spin states like $16^-$, $18^+$, and $19^+$, we need to perform shell model calculations by using neutron excitation across $N=126$.

\begin{figure}
	\centering
	\includegraphics[width=8.50cm,height=9cm]{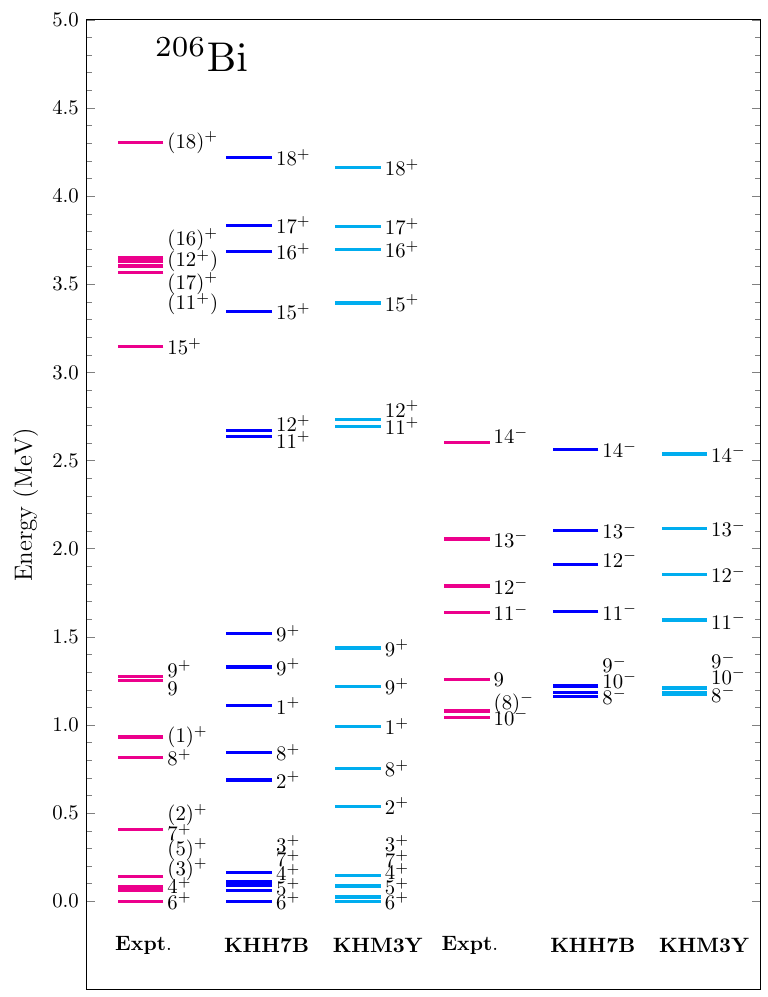}
	\caption{\label{206Bi} Comparison between the theoretical and experimental \cite{NNDC} energy levels in $^{206}$Bi isotope.}
\end{figure}

{\bf $^{206}$Bi}: The comparison between the theoretical and experimental energy spectra of $^{206}$Bi using the KHH7B and KHM3Y interactions is shown in Fig. \ref{206Bi}. For comparison, we have taken the experimental energy states up to 4.305 MeV. For the positive parity states, the lowest-lying state $6^+_{g.s.}$ is reproduced using both interactions. The yrast states $2^+_1$, $4^+_1$, and $6^+_1$ are obtained from the same dominant configuration i.e. $\pi(h_{9/2}^1)\otimes\nu (f_{5/2}^5p_{3/2}^4i_{13/2}^{14})$. These states are formed due to coupling of  $\pi(h_{9/2})$ and $\nu(f_{5/2})$ orbitals.
Experimentally, at 0.931 MeV, tentative spin state (1)$^+$ is obtained, whereas theoretically, the $1^+_1$ state lies 177 and 59 keV above using KHH7B and KHM3Y interaction, respectively. So, we may predict this state to be $1^+_1$. For the negative parity states experimentally, 10$^-$ is the lowest-lying state, while theoretically we get 8$^-$ as the lowest lying state with $\pi(h_{9/2}^1)$  $\otimes $  $\nu (f_{5/2}^6p_{3/2}^4i_{13/2}^{13})$ configuration and  probability 54.13 $\%$. Except for the first two states, i.e., $8^-$, and ${10}^-$, our calculations reproduce all the negative parity levels reasonably well with the same order, i.e., ${11}^-$--${12}^-$--${13^-}$--${14}^-$.   The $14^-_1\rightarrow13^-_1$ and $18^+_1\rightarrow17^+_1$ transitions occur via $E2+M1$ channel. Experimentally, the mixing ratios ($\delta(E2/M1)$) for $14^-_1\rightarrow13^-_1$ and $18^+_1\rightarrow17^+_1$ transitions were observed as -0.14(3) and -0.15(5), respectively \cite{Cieplicka}; the corresponding shell model $\delta(E2/M1)$ values are -0.09 and -0.05, both are consistent with the experimental data. The formula $\delta(E2/M1)=0.835E_{\gamma}$[MeV]$\frac{<E2>}{<M1>}[\frac{eb}{\mu_N/c}]$ is utilized in the present calculation, where $<E2>$, and $<M1>$ denote the shell model calculated $E2$ and $M1$ reduced matrix elements.  For the other transition $8^+_1\rightarrow 7^+_1$, the multipolarity mixing was suggested as $\delta \approx 0.5$ in the Ref. \cite{Scheveneels}. While the magnitude of the mixing ratio ($\delta(E2/M1)$) from the shell model calculation is obtained as 0.77.

\begin{figure}
	\centering
	\includegraphics[width=8.50cm,height=9cm]{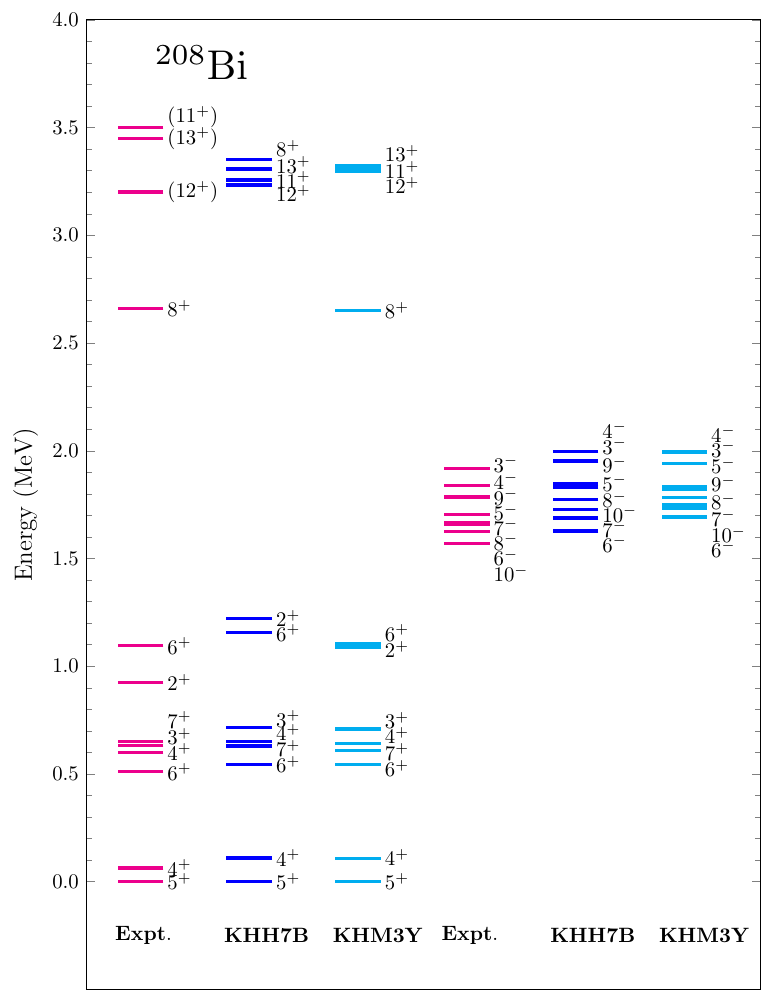}
	\caption{\label{208Bi} Comparison between the theoretical and experimental \cite{NNDC} energy levels in $^{208}$Bi isotope.}
\end{figure}

{\bf $^{208}$Bi}: Fig. \ref{208Bi} shows a comparison between the theoretical and experimental energy spectra of $^{208}$Bi using the KHH7B and KHM3Y interactions. We have considered the experimental energy states up to $\sim$ 3.500 MeV. 
With the KHH7B interaction the yrast states $4^+$, and $5^+$ are obtained from similar configurations i.e.  $\pi(h_{9/2}^1)\otimes\nu(p_{1/2}^{-1})$ with the probabilities 91.26 and 97.99 $\%$, respectively. Positive parity yrast states lying between 0.5 to 1.0 MeV are coming due to the coupling of $\pi(h_{9/2})$ and $\nu(f_{5/2})$ orbitals.  Our findings for the configuration of the above states support the previous study \cite{Alford}. Shell-model calculations using KHH7B interaction predict the $B(E2)$ values for $6^+_1\rightarrow4^+_1$ and $3^+_1\rightarrow5^+_1$ transitions as 0.1 and $0.12\times 10^{-2}$ W.u., while the corresponding experimental values are 0.10(3) and $0.25^{-8}_{+11}$ W.u., respectively. The structure change in the case of both transitions may cause the small $B(E2)$ values. The high-spin states above 2.5 MeV excitation energy are coming from the contribution of one valence proton particle and one valence neutron hole in the $\pi(i_{13/2})$ and $\nu(i_{13/2})$ orbitals, respectively. The positive parity yrast states up to 0.510 MeV excitation energy are well reproduced using both interactions, whereas the $8^+$ state at 2.660 MeV excitation energy is reproduced well by only KHM3Y interaction. The reason might be, the $8^+_1$ state using KHH7B interaction is formed due to $(\pi i_{13/2}^1\otimes \nu i_{13/2}^{-1})$ whereas according to Ref. \cite{crawley, Alford} configuration of this state should be $(\pi h_{9/2}^1 \otimes\nu f_{7/2}^{-1})$, since KHH7B model space does not consist $\nu (f_{7/2})$ orbital, whereas KHM3Y model space consists, therefore $8^+_1$ state [$(\pi h_{9/2}^1 \otimes\nu f_{7/2}^{-1})$] is reproduced very well by KHM3Y interaction. Experimentally, at 3.201 MeV, the tentative state $(12^+)$ is obtained; we can predict this state to be 12$^+_1$ by comparing our shell-model results. First, among the experimental negative parity states is 10$^-$ at 1.571 MeV excitation energy, whereas theoretically, we get 6$^-_1$ at 1.627 MeV using KHH7B interaction with the obtained configuration $\pi(i_{13/2}^1)\otimes\nu(f_{5/2}^6p_{3/2}^4p_{1/2}^1i_{13/2}^{14})$. By using the Nordheim rule also, we get the lowest lying negative parity state $6^-_1$, by the coupling of $(\pi i_{13/2}^1\otimes \nu p_{1/2}^1)$, maybe some tensor-force component is required to reproduce the experimentally observed lowest negative parity state. It is also notable that the energy of $10^-_1$ $[\pi(h_{9/2}^1)\otimes\nu(i_{13/2}^{-1})]$ state in the bismuth isotopes below $N=126$ drops steadily with the decrease of mass number experimentally and theoretically both (see Figs. \ref{204Bi}-\ref{208Bi}), which is due to the depletion in the contribution of $\nu(f_{5/2})$ and $\nu(p_{1/2})$ orbitals. Thus, the $\nu(i_{13/2})$ orbital is coming close to the Fermi surface.

\begin{figure}
	\centering
	\includegraphics[width=8.50cm,height=9cm]{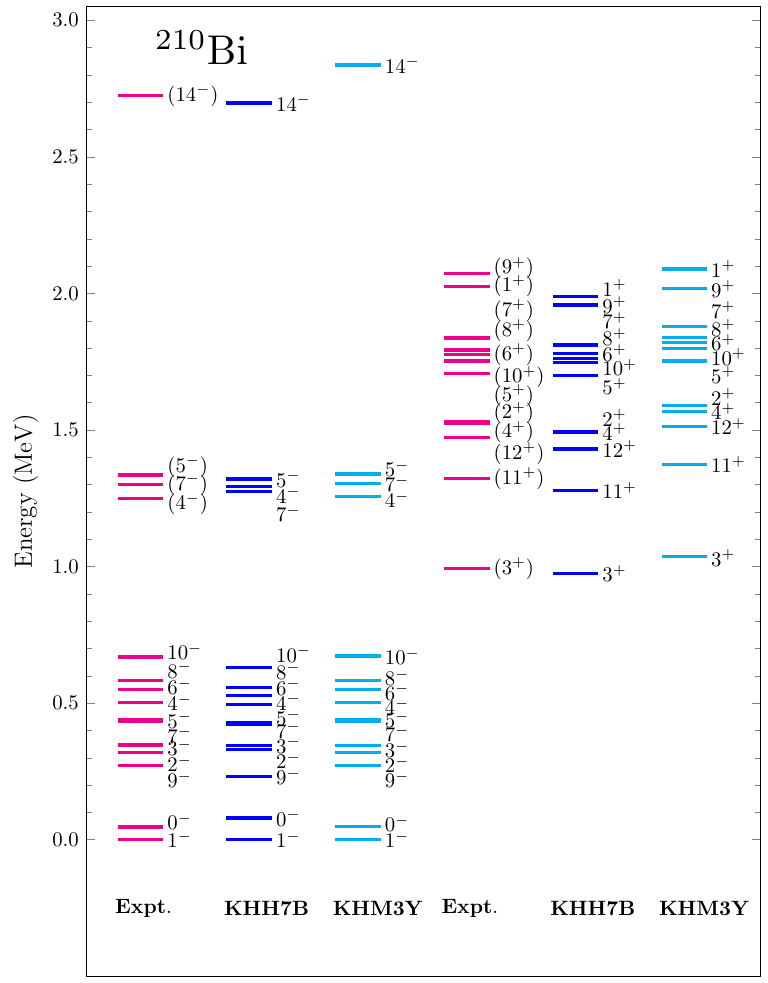}
	\caption{\label{210Bi} Comparison between the theoretical and experimental \cite{NNDC} energy levels in $^{210}$Bi isotope.}
\end{figure}

{\bf $^{210}$Bi}: Fig. \ref{210Bi} shows a comparison between the theoretical and experimental energy spectra of $^{210}$Bi using the KHH7B and KHM3Y interactions.  We have considered the experimental energy states up to 2.725 MeV. The $^{210}$Bi isotope provides a testing ground for the proton-neutron interaction of the shell model in reproducing the different energy states due to a single proton and neutron above $Z=82$ and $N=126$, respectively. Unlike other even-$A$ Bi isotopes, here, the lowest-lying state is the negative parity state. The yrast $0^--9^-$ states are favored by same dominant configuration $(\pi h_{9/2}^1\otimes\nu g_{9/2}^1)$ as suggested also in the Ref. \cite{Sheline}. Experimentally, 1$^-$ state is the lowest lying state, but according to Nordheim rule, due to coupling of $(\pi h_{9/2}^1\otimes\nu g_{9/2}^1)$, we should get 0$^-$ as the ground state, however, whole $0^-_1-9^-_1$ states are well reproduced by our calculation. The theoretical result for the 14$^-_1$ state supports the experimentally tentative state at 2.725 MeV excitation energy because, theoretically, this state lies at 2.697 and 2.836 MeV using KHH7B and KHM3Y interaction, respectively. Above 1.0 MeV, all the experimental states are tentative. Experimentally, for positive parity states, the lowest-lying state at 0.993 MeV excitation energy is $(3^+)$, our calculated $3^+_1$ state shows a small discrepancy in energy of 19 and 44 keV using KHH7B and KHM3Y interactions, respectively, and support this tentative state may be $3^+_1$.  For the positive and negative parity, all the experimentally tentative states are reproduced by the corresponding theoretical results.

\begin{figure}
	\centering
	\includegraphics[width=9.00cm,height=9cm]{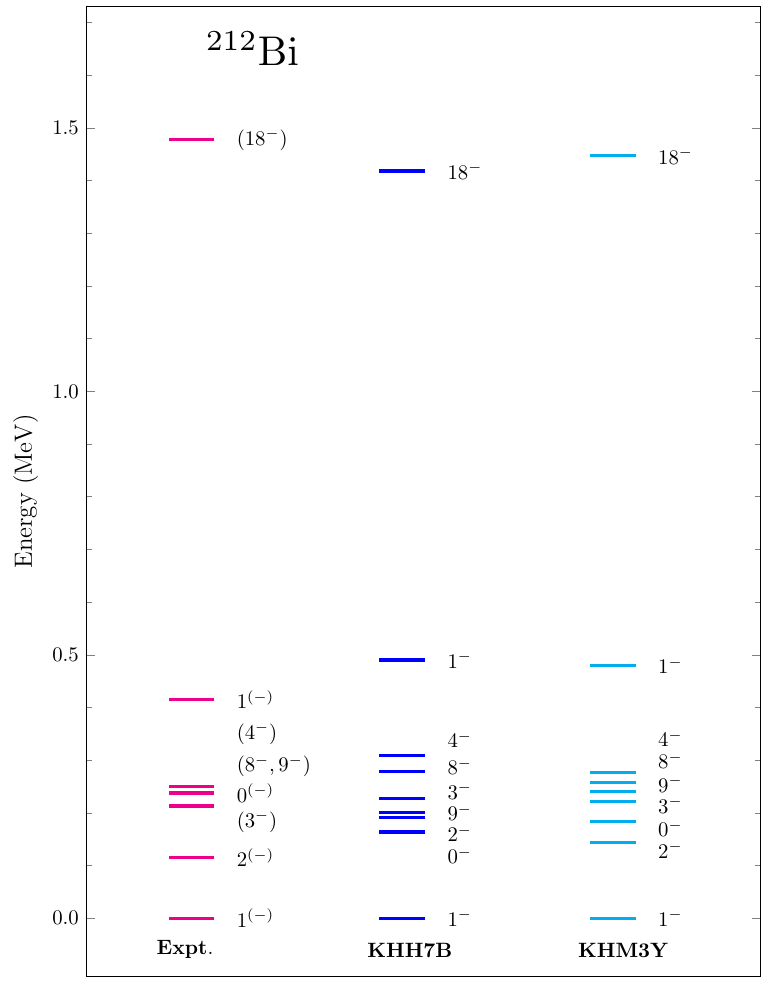}
	\caption{\label{212Bi} Comparison between the theoretical and experimental \cite{NNDC} energy levels in $^{212}$Bi isotope.}
\end{figure}

{\bf $^{212}$Bi}:  The theoretical and experimental energy spectra of $^{212}$Bi using the KHH7B and KHM3Y interactions are compared in Fig. \ref{212Bi}. We have considered the experimental energy states up to 1.5 MeV excitation energy and reported only those yrast and non-yrast SM states that correspond to the available experimental data. Experimentally, only negative parity states are available in this isotope; among these states, some have unconfirmed parity, and some have both unconfirmed spin and parity. The parity of the experimental ground state $1^{(-)}$ is tentative. Theoretically, we obtained $1^{-}$ as g.s. by both interactions and predicted this state should have negative parity. Our theoretical results corresponding to experimental states with tentative negative parity, such as $2^{(-)}$, $0^{(-)}$, and $1^{(-)}$ at 0.115, 0.238, and 0.415 MeV excitation energy are obtained at 0.144, 0.183, and 0.480 MeV, respectively using KHM3Y interaction which also supports the negative parity for these states. Experimentally, the lower limits of the $B(M1)$ transition from $1^-_2$ state to the $0^-_1$ and $1^-_{g.s}$ are $6.444\times 10^{-3}$, and $1.074\times 10^{-4}$ $\mu_N^2$, respectively \cite{NNDC_NUDAT}; the corresponding obtained values $1.847\times 10^{-2}$, and $1.306\times 10^{-3}$ $\mu_N^2$ from the KHH7B interaction, these values are consistent with the experimental data. Our SM results also support the experimentally tentative state $(18^-)$ at 1.478 MeV excitation energy, which corresponds to our theoretical $18^-_1$ state from both interactions. The yrast $0^-$, $1^-$, $2^-$, $3^-$, $4^-$, $8^-$, and $9^-$ states are formed from the domiant configuration [$\pi(h_{9/2}^1)\otimes\nu (g_{9/2}^3)$]. Although the positive parity state is not available experimentally, present KHH7B results predict the lowest-lying positive parity state is $3^+$ at 0.791 MeV. 

\begin{table*}
	\caption{\label{be2} The calculated (with KHH7B) $B(E2)$ values in units of W.u. for Bi isotopes compared to the experimental data (Expt.)
		\cite{NNDC,204BI,205BI,206BI,207BI,208BI,209BI,210BI,211BI,212BI,213BI} corresponding to $e_p$ = 1.5$e$ and $e_n$ = 0.5$e$.  }
	
	\begin{tabular}{rrc|cccc}
		\hline      
		& ${B(E2; J_i \rightarrow   J_f}$)  & \hspace{1.0cm}~~~~~ &  \hspace{1.0cm}~~~~~ ${B(E2; J_i \rightarrow   J_f}$) &      &\\
		\hline
		& &    &   &     &    \\
		$^{204}$Bi & Expt. & SM & $^{205}$Bi  & Expt. & SM     \\
		\hline
		{5}$^+_1$ $\rightarrow$ {6}$^+_1$&NA &2.1  & 7/2$^-_1$ $\rightarrow$ 9/2$^-_1$  & NA    & 0.4 \\
		{4}$^+_1$ $\rightarrow$ {6}$^+_1$&NA &0.5    & 11/2$^-_1$ $\rightarrow$ 9/2$^-_1$  & NA    & 2.0   \\
		{4}$^+_1$ $\rightarrow$ {5}$^+_1$&NA &2.7    &   25/2$^+_1$ $\rightarrow$ 21/2$^+_1$ &0.59(8) &0.7 \\
		\hline
		& &    &   &     &    \\
		$^{206}$Bi  & Expt. & SM  & $^{207}$Bi & Expt. & SM  \\
		& &    &   &     &    \\
		\hline
		{4}$^+_1$ $\rightarrow$ {6}$^+_1$ &0.0180(6) &5.91$\times$10$^{-6}$ & 11/2$^-_1$ $\rightarrow$ 9/2$^-_1$   &  NA   &1.8    \\
		{3}$^+_1$ $\rightarrow$ {4}$^+_1$&NA &2.4    & 7/2$^-_1$ $\rightarrow$ 9/2$^-_1$  &  NA   & 0.4   \\
		{5}$^+_1$ $\rightarrow$ {4}$^+_1$&NA &2.6    & 7/2$^-_1$ $\rightarrow$ 11/2$^-_1$  & NA    & 0.1   \\
		{5}$^+_1$ $\rightarrow$ {6}$^+_1$&NA &3.2    &  13/2$^-_1$ $\rightarrow$ 9/2$^-_1$  & NA    &  0.6  \\

		\hline
		& &    &   &     &    \\
		$^{208}$Bi  & Expt. & SM  & $^{209}$Bi & Expt. & SM  \\
		& &    &   &     &    \\
		\hline
		6$^+_1$ $\rightarrow$ 4$^+_1$ &0.10(3) &0.1 & 3/2$^-_1$ $\rightarrow$ 7/2$^-_1$ &7(5) &2.54$\times$10$^{-4}$\\
		6$^+_1$ $\rightarrow$ 5$^+_1$ &1.5(5)&0.6 &5/2$^-_1$ $\rightarrow$ 9/2$^-_1$ &4.4(6) &1.36$\times$10$^{-5}$\\
		7$^+_1$ $\rightarrow$ 5$^+_1$ & $\leq$0.017&0.3 &7/2$^-_1$ $\rightarrow$ 9/2$^-_1$ &0.44(9) &0.3\\
		3$^+_1$ $\rightarrow$ 5$^+_1$ &0.25$_{+11}^{-8}$ &0.12$\times$10$^{-2}$ & $19/2^+_1$ $\rightarrow$ $15/2^+_1$  & 0.387(12)& 0.2 \\
		{4}$^+_1$ $\rightarrow$ {5}$^+_1$&NA &0.4 &  $15/2^+_1$ $\rightarrow$ $13/2^+_1$ &  0.0025(15)   & 6.09$\times 10^{-7}$   \\
		\hline
		& &    &   &     &    \\
		$^{210}$Bi  & Expt. & SM  & $^{211}$Bi & Expt. & SM  \\
		& &    &   &     &    \\
		\hline
		{3}$^-_1$ $\rightarrow$ {1}$^-_1$&NA &2.0  & 7/2$^-_1$ $\rightarrow$ 9/2$^-_1$ &1.07(10)& 0.2\\
		{3}$^-_1$ $\rightarrow$ {2}$^-_1$&NA &0.2  &  21/2$^-_1$ $\rightarrow$ 17/2$^-_1$ &1.44(11)&0.7\\
		{6}$^-_1$ $\rightarrow$ {4}$^-_1$ &   NA  & 1.4 &9/2$^-_3$ $\rightarrow$ 7/2$^-_1$ &$>$ 0.00015&0.1\\
		{8}$^-_1$ $\rightarrow$ {6}$^-_1$  &  NA   & 0.5 & 9/2$^-_3$ $\rightarrow$ 9/2$^-_1$ &$>$ 0.0031&4.1$\times 10^{-2}$ \\
		
		\hline
		& &    &   &     &    \\
		$^{212}$Bi  & Expt. & SM  & $^{213}$Bi & Expt. & SM  \\
		& &    &   &     &    \\
		\hline
		3$^-_1$ $\rightarrow$ 2$^-_1$       & NA& 0.7  &  {7/2}$^-_1$ $\rightarrow$ {9/2}$^-_1$ &NA &0.2\\
		4$^-_1$ $\rightarrow$ 2$^-_1$&NA & 2.2 &  & &\\
		
		\hline
		
	\end{tabular}
\end{table*}

\subsection{\label{odd-odd} Results for odd-$A$ $^{205-213}$Bi isotopes}

\begin{figure}
	\centering
	\includegraphics[width=8.5cm,height=9cm]{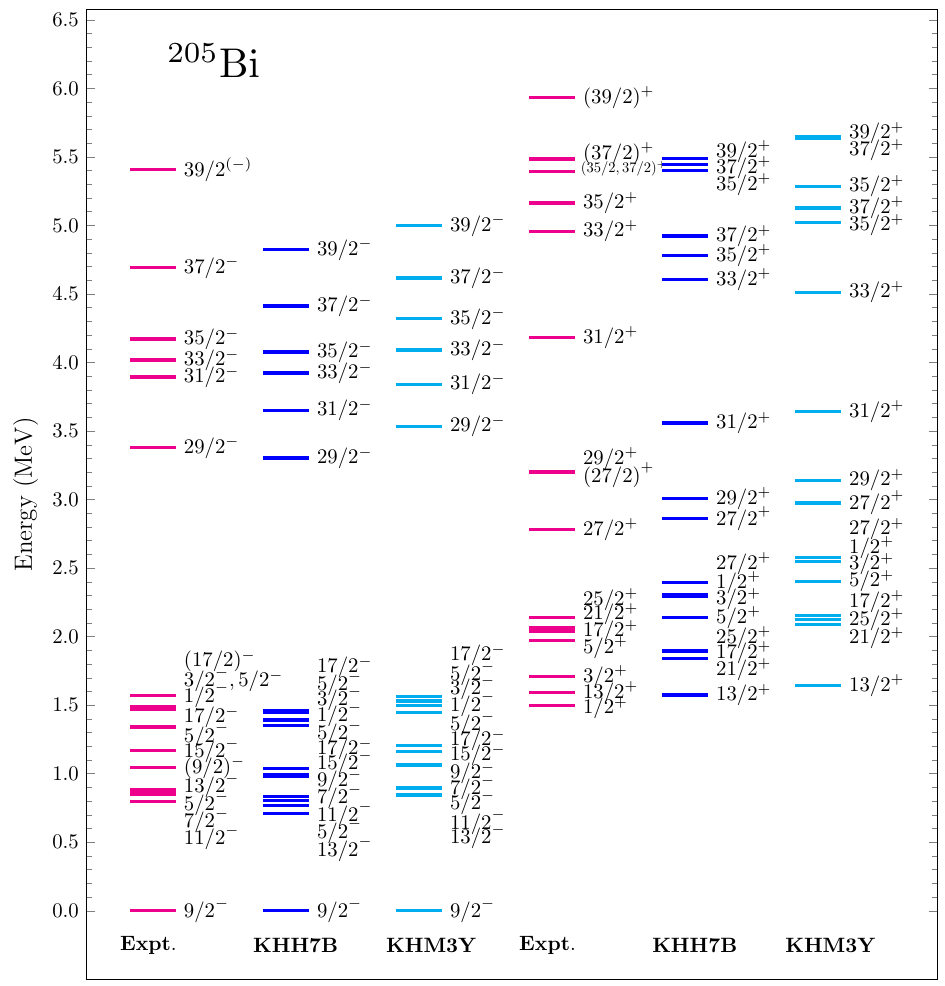}
	\caption{\label{205Bi} Comparison between the theoretical and experimental \cite{NNDC} energy levels in $^{205}$Bi isotope.}
\end{figure}

{\bf $^{205}$Bi}:  Fig. \ref{205Bi} shows a comparison between the theoretical and experimental energy spectra of $^{205}$Bi utilizing the KHH7B and KHM3Y interactions.  We have considered the experimental energy states up to $\sim$ 6.0 MeV. We are able to reproduce the 9/2$^-_{g.s.}$ correctly by both interactions. Shell-model predicted negative parity states above 1.0 MeV excitation energy are in the same order as in the experimental spectra. The configuration of yrast states $1/2^-$ to $17/2^-$ obtained by KHH7B interaction is $\pi(h_{9/2}^1)\otimes\nu(f_{5/2}^4p_{3/2}^4i_{13/2}^{14})$, i.e., these yrast states are formed due to coupling of one proton particle in $h_{9/2}$ and two neutron holes in $f_{5/2}$ orbital. Among these yrast states, experimentally, the nature of $17/2^-\rightarrow15/2^-$ transition should be $M1$ type \cite{brock, T.Lonnroth}. Although the experimental $B(M1)$ value is not available yet, the shell model obtained $B(M1;17/2^-_1\rightarrow15/2^-_1)$ transition is 0.055 $\mu_N^2$; which might be helpful for the comparison with the upcoming experimental value. The excited states above 4.5 MeV are under-predicted from the KHH7B interaction in comparison to the experimental data in both negative and positive parity states. Thus, we can suggest that the obtained $39/2^-_1$ state from both interactions corresponds to the experimentally tentative state $39/2^{(-)}$. It may be possible that these high-spin states arise due to the core excitation in the neutron shells across $N=126$. Experimentally, the lowest positive parity state is 1/2$^+$ at 1.497 MeV, whereas from both interactions, the lowest lying state is 13/2$^+$. The $13/2^+_1$ state obtained from configuration $\pi(i_{13/2}^1)\otimes\nu(f_{5/2}^4p_{3/2}^4i_{13/2}^{14})$, shows single-particle nature, i.e. formed purely by $\pi i_{13/2}$ orbital.  Previously \cite{Hopke} it was suggested that the $1/2^+_1$ state is characterized by proton-aligned configuration $\pi(h_{9/2}^2s_{1/2}^{-1})$. But in our present calculation, we have not included $\pi(s_{1/2})$ orbital below $Z=82$. This might be a possible cause for the overprediction of the $1/2^+_1$ state. Our theoretical result suggests the tentative state $(27/2)^+$ and $(39/2)^+$ at 3.198 and 5.931 MeV excitation energy to be $27/2^+_2$, and $39/2^+_1$, respectively, these predictions are based on the order of calculated spectra.


\begin{figure}
	\centering
	\includegraphics[width=8.5cm,height=9cm]{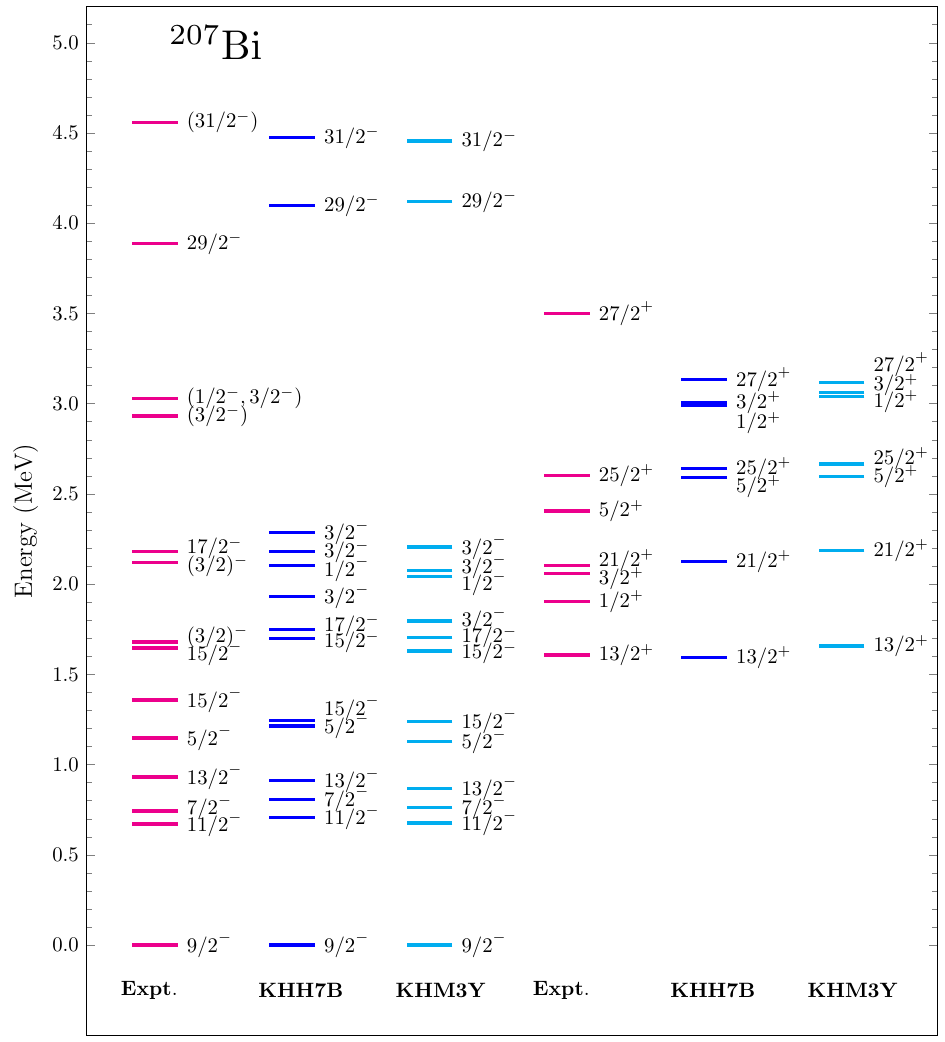}
	\caption{\label{207Bi} Comparison between the theoretical and experimental \cite{NNDC} energy levels in $^{207}$Bi isotope.}
\end{figure}

{\bf $^{207}$Bi}: Fig. \ref{207Bi} shows a comparison between the theoretical and experimental energy spectra of $^{207}$Bi using the KHH7B and KHM3Y interactions. We have considered only those yrast and non-yrast states which correspond to the available experimental data. We are able to reproduce the ground state 9/2$^-$ very well. Upto 1.148 MeV excitation energy, both the interactions reproduce spectra reasonably well for negative parity states. The yrast states $3/2^-$ to $15/2^-$ except $9/2^-$ are coming from the dominant $\pi(h_{9/2}^1)\otimes\nu(f_{5/2}^5p_{3/2}^4p_{1/2}^1i_{13/2}^{14})$ configuration with KHH7B interaction, whereas configuration of $9/2^-_1$ state is $\pi(h_{9/2}^1)$  $\otimes $  $\nu (f_{5/2}^6p_{3/2}^4i_{13/2}^{14})$, obtained purely by $\pi(h_{9/2})$ orbital thus the $9/2^-_1$ state shows single particle nature.  The experimentally observed levels in $^{207}$Bi were interpreted as resulting from the coupling of two neutron-hole states in $^{206}$Pb with a valence proton as reported in Ref. \cite{bergstrom}, our SM results support the configurations. Our theoretical calculation suggests the tentative experimental level (31/2$^-$) at 4.559 MeV excitation energy to be $31/2^-_1$. For the positive parity state, experimentally, the 13/2$^+$ is the lowest-lying state, which is reproduced very well from both interactions. Our theoretical calculation shows good agreement for the positive parity states up to 2.601 MeV except for 1/2$^+$ and 3/2$^+$ states. As discussed in the case of the previous isotope, the configuration of $1/2^+_1$ state in bismuth isotopes was suggested as $\pi(h_{9/2}^2s_{1/2}^{-1})$ in Ref. \cite{Hopke}. Similarly, in $^{207}$Bi, it was suggested that the $3/2^+_1$ state is favored by $[\pi(h_{9/2}^2d_{3/2}^{-1})\otimes \nu(p_{1/2}^{-2})]$ configuration \cite{Astner}. In both cases, a proton hole in the orbital below $Z=82$ shell-closure dominates. But, these $\pi(s_{1/2})$ and $\pi(d_{3/2})$ orbitals are not included in our model space. It might be a possible reason for the overprediction of both states. Results corresponding to 21/2$^+$ and 25/2$^+$ are nicely reproduced using both interactions.

\begin{figure}
	\centering
	\includegraphics[width=8.5cm,height=9cm]{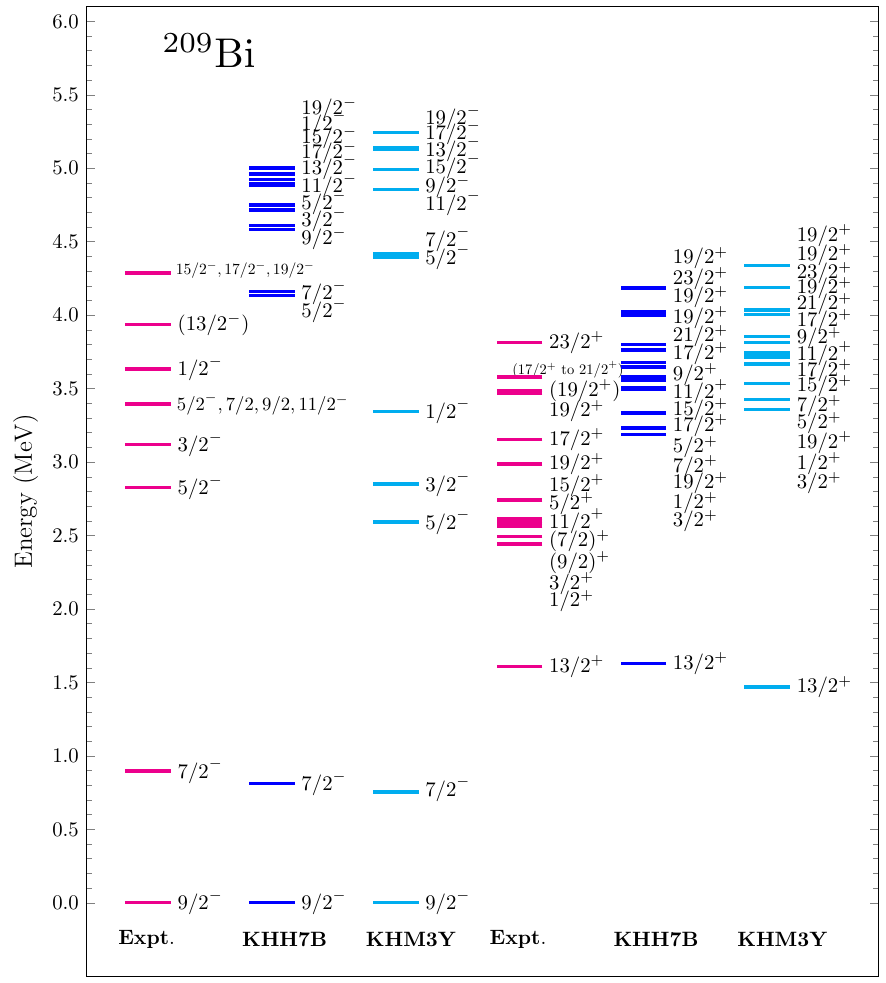}
	\caption{\label{209Bi} Comparison between the theoretical and experimental \cite{NNDC} energy levels in $^{209}$Bi isotope.}
\end{figure}


{\bf $^{209}$Bi}: Fig. \ref{209Bi} shows a comparison of the theoretical and experimental energy levels in $^{209}$Bi using the KHH7B and KHM3Y interactions. We have considered the experimental energy states up to $\sim$ 4.3 MeV. We are able to calculate only yrast states up to $13/2^+$ using KHH7B interaction without core-excitation due to completely filled neutron orbitals below $N=126$. In order to get states above $13/2^+$, we have performed $1p-1h$ excitation across neutron orbitals above $N=126$. We are able to reproduce negative parity yrast states reasonably well up to 3.2 MeV excitation energy using KHM3Y interaction. Our theoretical results using the KHH7B interaction are over-predicted for both the parities above 1.608 MeV excitation energy. Except the single-particle states $9/2^-_{g.s.}$, $7/2^-$, and $13/2^+$, all the excited states are coming from $2p-1h$ excitation from the KHH7B interaction. All the states above 1.608 MeV except the yrast states $1/2^-$, $3/2^-$, and $5/2^-$ are over-predicted using KHM3Y interaction. Experimentally, at 3.579 MeV, tentative state ($17/2^+$ to $21/2^+$) is observed, whereas theoretically, $17/2^+_2$ state lies at 3.677, and 3.853 MeV using KHH7B and KHM3Y interaction, respectively which is close to its experimentally tentative level, so we can predict this state to be $17/2^+_2$. Experimentally, the branching ratios ($B.R.$) for the $5/2^-_1\rightarrow 9/2^-_{g.s.}$ and $5/2^-_1\rightarrow 7/2^-_1$ $E2$ transitions were proposed as 76(4)\% and 24(6)\%, respectively \cite{Maier}. Although our calculation shows weak $B(E2;5/2^-_1\rightarrow 9/2^-_{g.s.})$ transition compared to the experimental value. But, our calculated $B.R.$ for $5/2^-_1\rightarrow 9/2^-_{g.s.}$ (69.64\%) and $5/2^-_1\rightarrow 7/2^-_1$ (30.36\%) transitions do not much differ from the experimental data. The $B.R.$ are estimated by the formula, $B.R.^{(k)}=t_{1/2}^{total}/t_{1/2}^{(k)}$; where, $t_{1/2}^{total}$ and $t_{1/2}^{(k)}$ stand for the total and partial half-life of the decay, respectively, and $k$ denote the final states of the decay.  Similarly, as reported in Ref. \cite{Maier}, the $17/2^+_1$ state decays via $17/2^+_1\rightarrow 15/2^+_1$ (93.5(5)\%) and $17/2^+_1\rightarrow 19/2^+_1$ (6.3(3)\%) branching with $M1+E2$ and $M1$ transitions, respectively. Our obtained values of $B.R.$ with SM for $17/2^+_1\rightarrow 15/2^+_1$ (77.8\%) and $17/2^+_1\rightarrow 19/2^+_1$ (22.2\%) transitions show small deviation from the experimental values.  The $B(E2)$ transitions are forbidden between the $3/2^-_1$ [$\pi(i_{13/2}^1)\otimes \nu(p_{1/2}^1 g_{9/2}^1)$] and $7/2^-_1$ [$\pi(f_{7/2}^1)$] state according to their congifuration. The similar case also happens for the transition between $15/2^+_1$ [$\pi(h_{9/2}^1)\otimes \nu(p_{1/2}^1 g_{9/2}^1)$] and $13/2^+_1$ [$\pi(i_{13/2}^1)$] state. The $B(E2)$ transition between $5/2^-_1$ [$\pi(h_{9/2}^1)\otimes \nu(p_{1/2}^1 j_{15/2}^1)$] and $9/2^-_1$ [$\pi(h_{9/2}^1)$] state is not completely forbidden. But, the structure change might be a reason for the small $B(E2)$ value from the shell-model.



\begin{figure}
\centering
\includegraphics[width=8.5cm,height=9cm]{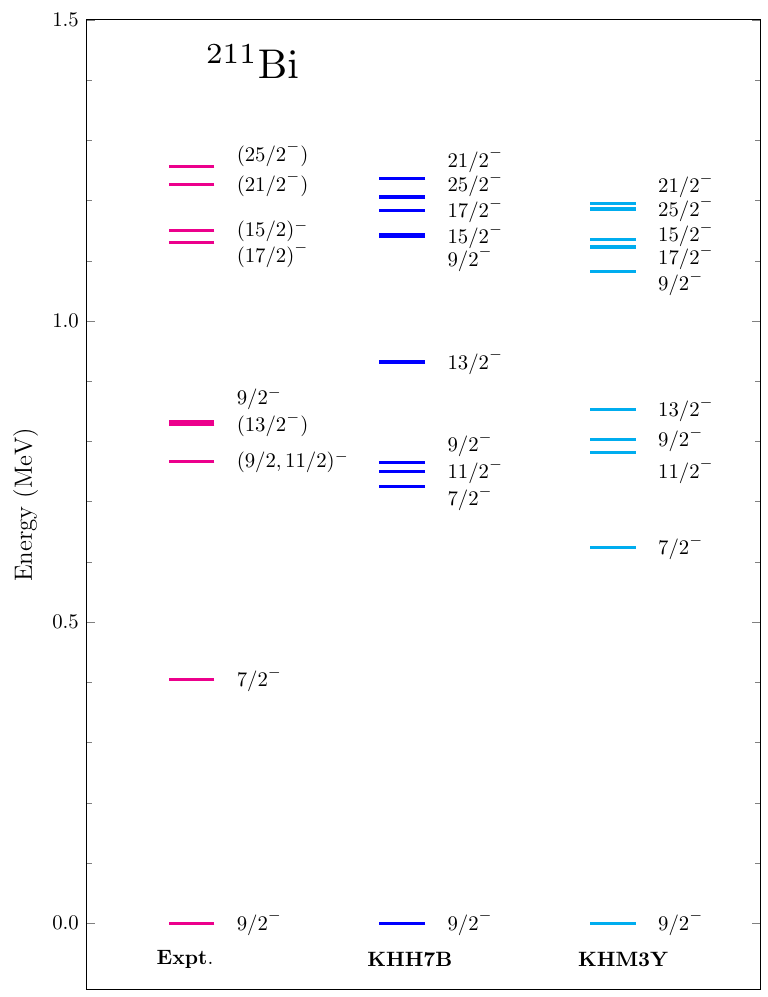}
\caption{\label{211Bi} Comparison between the theoretical and experimental \cite{NNDC} energy levels in $^{211}$Bi isotope.}
\end{figure}

{\bf $^{211}$Bi}: Fig. \ref{211Bi} shows a comparison between the theoretical and experimental energy spectra of $^{211}$Bi using the KHH7B and KHM3Y interactions. We have considered the experimental energy states up to 1.257 MeV excitation energy; up to this energy, only negative parity states are available for this isotope. Our calculations reproduce the ground state, i.e., $9/2^-$ state, correctly by both interactions. The configuration of the $9/2^-_1$ state obtained from the KHH7B interaction is $\pi(h_{9/2}^1)\otimes\nu(g_{9/2}^2)$ with probability 64.81 $\%$, thus $9/2^-_1$ state shows single-particle nature, which is formed due to the contribution of a valence proton in $\pi(h_{9/2})$ orbital. By using KHH7B interaction, all the states reported here except $7/2^-$ are coming due to $\pi (h_{9/2}^1)\otimes\nu (g_{9/2}^2)$. The calculated $7/2^-_1$ [$\pi (f_{7/2}^1) \otimes \nu(g_{9/2}^2)$] state is found at very high energy, in Ref. \cite{waldridge} it is shown that this state is coming due to coupling of single proton state $f_{7/2}$ and the $7/2^-$ member of the $ \pi (i_{13/2}) \otimes ^{208}$Pb$(3^-)$ core multiplet, thus core excitation is needed for the description of this state. The predicted $B(E2)$ values in $^{211}$Bi are not much deviating from the experimental data as compared in the case of $^{209}$Bi. Shell-model calculations suggest larger $B(E2)$ values for $9/2^-_3\rightarrow7/2^-_1$ and $9/2^-_3\rightarrow 9/2^-_1$ transitions compared to the lower limits of experimental data. However, a relatively smaller $B(E2;7/2^-_1\rightarrow 9/2^-_1)$ value from the shell-model might be due to the different structures of both states. All the states above 0.75 MeV are tentative. Experimentally at 0.766 MeV, we have $(9/2,11/2)^-$ state, whereas theoretically, 9/2$^-_2$ and 11/2$^-_1$ states lie below 1 and 16 keV using KHH7B interaction, respectively. There is another experimental $9/2^-$ state (at 0.832 MeV) in the same region, thus, we propose the state at 0.766 MeV may be 11/2$^-_1$.


\begin{figure}
\centering
\includegraphics[width=9.00cm,height=9cm]{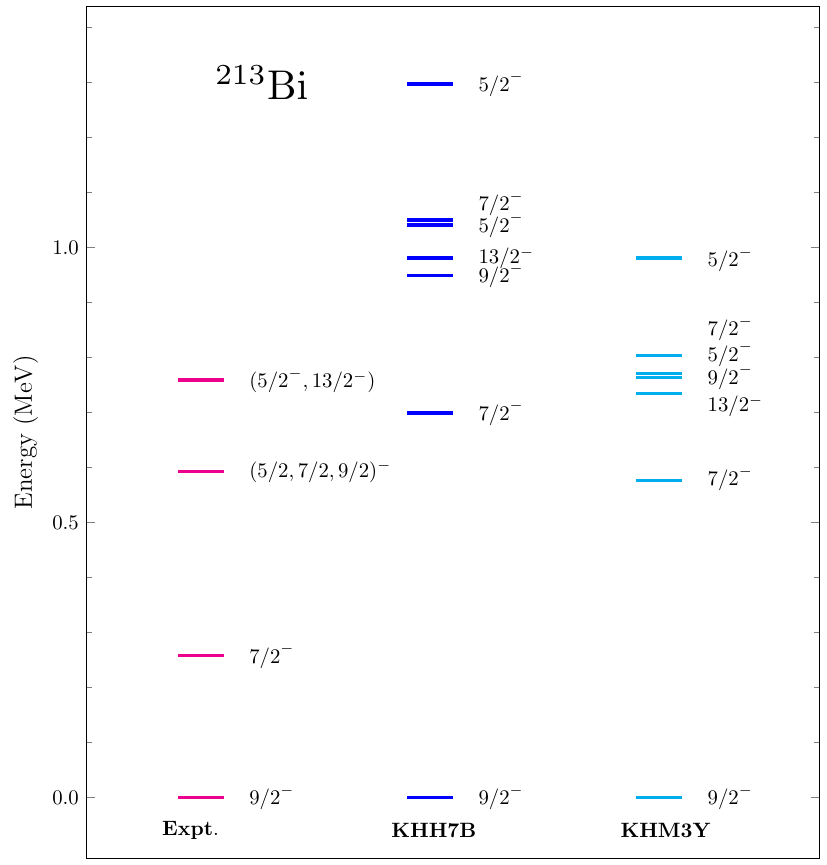}
\caption{\label{213Bi} Comparison between the theoretical and experimental \cite{NNDC} energy levels in $^{213}$Bi isotope.}
\end{figure}

{\bf $^{213}$Bi}: Fig. \ref{213Bi} shows a comparison between the theoretical and experimental energy spectra of $^{213}$Bi using the KHH7B and KHM3Y interactions. We have considered the experimental energy states up to $\sim$ 0.8 MeV, and reported only the yrast and non-yrast shell-model states that correspond to the available experimental data. In this isotope, only $9/2^-_1$ and $7/2^-_1$ states are experimentally confirmed. The 9/2$^-_{g.s.}$ is reproduced by both interactions nicely. The dominant configuration of the ground state obtained using KHH7B interaction is $\pi(h_{9/2}^1)\otimes\nu(i_{11/2}^2g_{9/2}^2)$ with 36.71\% probability. Yrast and non-yrast $5/2^-$ and yrast $13/2^-$ states are coming from $\pi (h_{9/2}^1)\otimes\nu (g_{9/2}^4)$, whereas $7/2^-_1$ state is coming due to $\pi (f_{7/2}^1)\otimes\nu (g_{9/2}^4)$. Similar to $^{211}$Bi, the $7/2^-_1$ state lies at very high energy in our calculation for $^{213}$Bi also. Experimentally, at 0.593 MeV (5/2,7/2,9/2)$^-$ state shows degeneracy, whereas corresponding theoretically obtained levels are non-degenerate. Except for the ground state, all other states show large discrepancies in their excitation energy from SM. The difference in energy, corresponding to the experimental levels, is less  using KHM3Y  interaction than KHH7B  interaction,  which is due to the large model space available in KHM3Y interaction; this demonstrates we need to enlarge model space to reproduce experimental data. Experimentally, at 0.758 MeV, the tentative state is ($5/2^-$, $13/2^-$). Using KHM3Y interaction, the $13/2^-_1$ state is at 0.733 MeV; therefore, we can predict the state at 0.758 MeV to be $13/2^-_1$.

\begin{figure*}
\includegraphics[height=5.0cm, width=5.33cm]{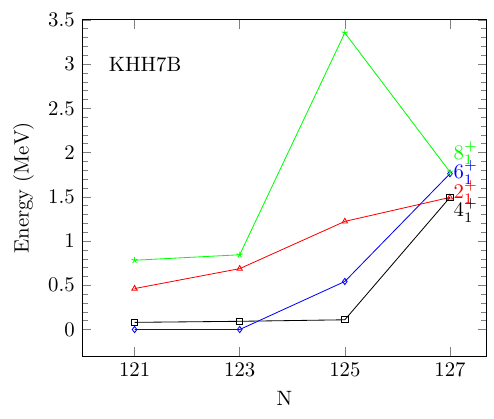}
\includegraphics[height=5.0cm, width=5.33cm]{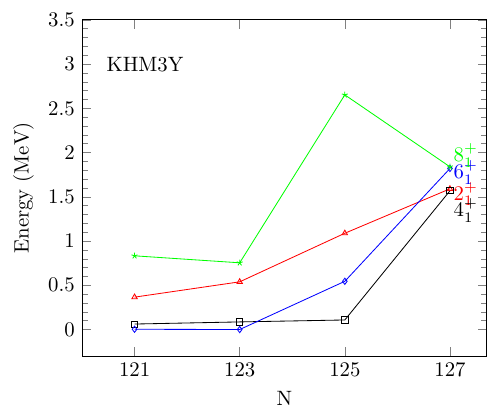}
\includegraphics[height=5.0cm, width=5.33cm]{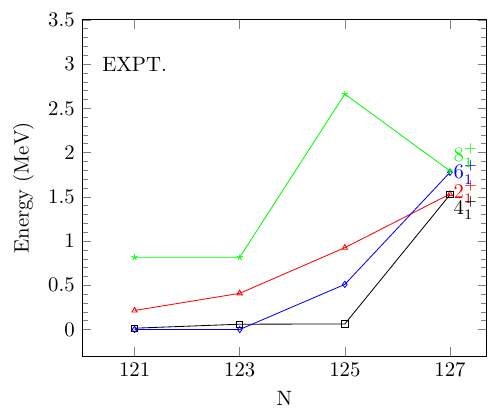}
\includegraphics[height=5.0cm, width=5.33cm]{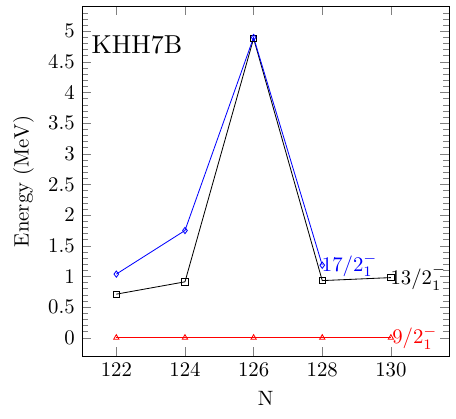}
\includegraphics[height=5.0cm, width=5.33cm]{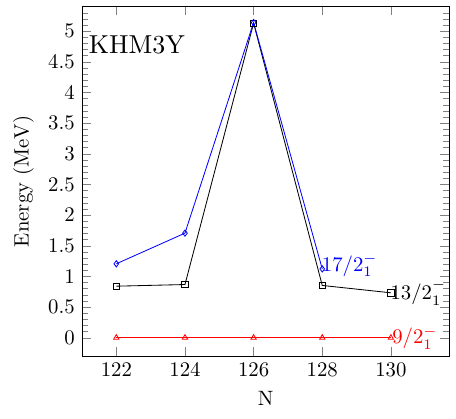}
\includegraphics[height=5.0cm, width=5.33cm]{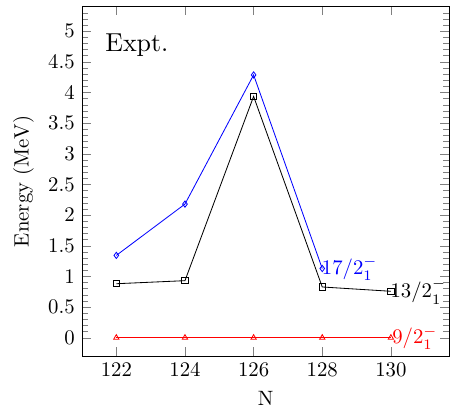}
\caption{\label{systematics_even} Energy level systematic of even-$A$ (upper panel) and odd-$A$ (lower panel) Bi isotopes.}

\end{figure*}

The results yielded by both KHH7B and KHM3Y interactions show good agreement with the experimental data. We can see as we go beyond $^{210}$Bi KHM3Y interaction gives better results than KHH7B interaction due to the enlarged model space of KHM3Y interaction. We have shown energy level systematics for even $2^+_1$ - $8^+_1$ states and odd $9/2^-_1$, $13/2^-_1$, and $17/2^-_1$ states in Fig. \ref{systematics_even}. Both comparisons show reasonable similarity between theoretical and experimental data.  All the low-spin states ($2^+-6^+$) in $^{208}$Bi are coming with the contribution of single-proton in $h_{9/2}$ orbital, and any of the neutron configuration between the $\nu(f_{5/2}^{-1})$, $\nu(p_{1/2}^{-1})$, and $\nu(p_{3/2}^{-1})$. However, as discussed in the earlier section, the $8^+_1$ state is primarly stems from either $\pi(i_{13/2}^1)\otimes \nu(i_{13/2}^{-1})$ (with KHH7B) or $\pi(h_{9/2}^1)\otimes \nu(f_{7/2}^{-1})$ (with KHM3Y). Here, the $\nu(f_{7/2})$ orbital keeps the largest single-particle energy gap from the Fermi level compared to the $\nu(p_{1/2})$, $\nu(p_{3/2})$, and $\nu{i_{13/2}}$ orbitals. Similarly, the $\pi(i_{13/2})$ and $\nu{i_{13/2}}$ orbitals (in KHH7B) also keep the largest single-particle energy gap from the Fermi level compared to any other involved proton and neutron orbital, respectively. Thus, we may conclude that the $8^+_1$ states from both the KHH7B and KHM3Y interaction originate due to the contribution of higher-lying proton and neutron orbitals. It may be a possible cause for the large energy gap between the low-spin states ($2^+-6^+$) and $8^+_1$ state in $^{208}$Bi.

\subsection{\label{EM}Magnetic and Quadrupole Moments}

\begin{table}        
	\begin{center}
		
		\caption{\label{qm} Comparison between the theoretical (with KHH7B) and experimental \cite{NNDC,204BI,205BI,206BI,207BI,208BI,209BI,210BI,211BI,212BI,213BI} magnetic dipole moments $\mu$ (in $\mu_N$) and electric quadrupole moments $Q$ (in $eb$) in Bi isotopes. We have taken the effective charges as $(e_p,e_n)=$ (1.5, 0.5)$e$.  The  gyromagnetic ratios for orbital angular momenta are taken as $g_l^\nu$ = 0.00, $g_l^\pi$ = 1.00, and for spin angular momenta as $g_s^\nu$ = -2.678, $g_s^\pi$ = 3.910.}
		
		\begin{tabular}{c  c c  c   c c c c}
			\hline
			A    & $J^\pi$   & $\mu$ ($\mu_N$) &  SM   & $Q (eb) $   &  SM \\
			\hline
			$^{204}$Bi & $6^+_1$  &  +4.322(15)      &  3.70   &      -0.49(15)       &  -0.48  \\
			& $10^-_1$  &  2.36(23)      &  1.67   &     0.063(12)       &  -0.05  \\
			& $4^+_1$  &  NA      &  2.91   &     NA      &  -0.31  \\
			& $5^+_1$  &  NA     &  3.34   &     NA       &  -0.36  \\
			
			$^{205}$Bi & $9/2^-_1$  &+4.065(7)   & 3.29    &  -0.81(3)           & -0.55  \\
			& $21/2^+_1$  &2.70(4) &  1.64  &       NA     & 0.30  \\
			& $25/2^+_1$  & 3.21(5)& 2.56   &        NA    & 0.09  \\
			& $7/2^-_1$  & NA& 2.73   &     NA       & -0.20  \\
			& $11/2^-_1$  & NA& 3.28   &      NA      & -0.38  \\
			& $13/2^-_1$  & NA& 3.79   &       NA     & -0.60  \\

			$^{206}$Bi & $6^+_1$  &+4.361(8) & 3.73   &  -0.39(4)          & -0.40  \\
			& $10^-_1$  &   2.644(14)      &  1.69  &     0.049(9)       & -0.07  \\
			& $4^+_1$  & NA& 2.93   &     NA       & -0.34  \\
			& $5^+_1$  & NA& 3.36   &      NA      & -0.34  \\

			$^{207}$Bi & $9/2^-_1$  & +4.0915(9)& 3.32  & -0.545(38)           & -0.53  \\
			& $21/2^+_1$  &+3.41(6) & 2.37   &  0.044(8)          &  -0.05 \\
			& $7/2^-_1$  & NA& 2.85   &     NA       & -0.38  \\
			& $11/2^-_1$  & NA& 3.23   &      NA      & -0.37  \\
			& $13/2^-_1$  & NA& 3.63   &      NA      & -0.31  \\

			$^{208}$Bi & $5^+_1$  & +4.578(13)&  3.64  &   -0.51(7)         & -0.45  \\
			& $10^-_1$  &2.672(14) &  1.63  &      NA      & -0.10  \\
			& $4^+_1$  &NA &  3.32  &      NA      & -0.45  \\
			& $6^+_1$  &NA &  3.70  &       NA     & -0.32  \\

			$^{209}$Bi  & $9/2^-_1$  &+4.1103(5) &  3.38  &    -0.55(1)        & -0.43  \\
			& $9/2^+_1$  &3.5(7) &  1.46  &  0.11(5)          & 0.16  \\
			& $13/2^+_1$  &NA &  7.64  &  -0.37(3)          & -0.55  \\
			& $15/2^+_1$  &6.2(12) & 2.06   &    0.0(4)        & -0.23  \\
			& $19/2^+_1$  &3.50(8) &  2.37  &      NA      & -0.62  \\
			$^{210}$Bi & $1^-_1$  & -0.04451(6)& 0.17   &+0.136(1)            & 0.15  \\
			& $9^-_1$  &2.728(42) &  2.00  &-0.471(59)            & -0.59  \\
			& $5^-_1$  &+1.530(45) &  1.10  &      NA      &  0.04 \\
			& $7^-_1$  &+2.114(49) &   1.55 &      NA      & -0.22  \\

			$^{211}$Bi  & $9/2^-_1$  & (+)3.79(7)&   3.21 &     -0.7(3)       & -0.50  \\
			& $7/2^-_1$  & +4.5(7)& 3.88   &       NA     & -0.42  \\
			& $13/2^-_1$  & NA& 2.31   &    NA        & -0.31  \\
			
			\hline

\end{tabular}
\end{center}
\end{table}

\begin{table}        
	\begin{center}
		
		{\bf Table 2.} (Continued)
		
		\begin{tabular}{c  c c  c   c c c c}
			\hline
			A    & $J^\pi$   & $\mu$ ($\mu_N$) &  SM   & $Q (eb) $   &  SM \\
			\hline

			$^{212}$Bi  & $1^{-}_1$  &+0.32(4) &  0.28  &   +0.1(4)         & 0.13  \\
			& $2^{-}_1$  & NA& 0.66   &    NA        & 0.18  \\
			&$3^-_1$  & NA& 0.73   &    NA        & 0.19  \\

			$^{213}$Bi  & $9/2^-_1$  &+3.699(7) & 3.16   &  -0.83(5)          &  -0.56 \\
			& $7/2^-_1$  &NA & 4.31  &  NA          &  -0.48 \\
			\hline
			
		\end{tabular}
	\end{center}
\end{table}

\begin{figure*}
\includegraphics[width=8.37cm]{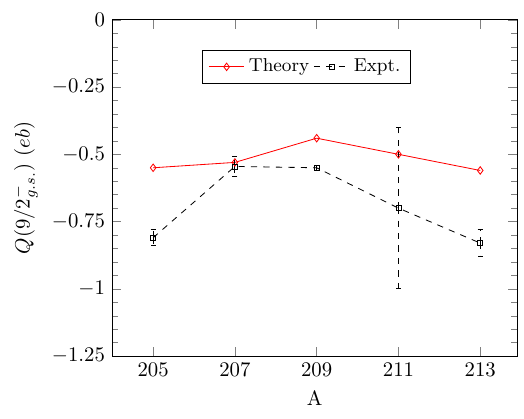}
\includegraphics[width=8.00cm]{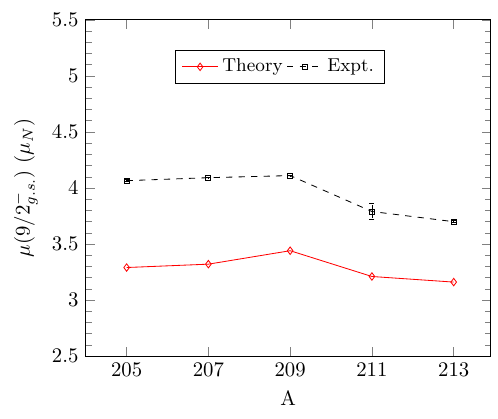}
\caption{\label{quad_magnetic} Comparison between the theoretical and experimental electric quadrupole (left) and magnetic dipole moment (right) of $9/2^-_{g.s.}$ for odd-$A$ $^{205-213}$Bi isotopes.}
\end{figure*}

In this section, we discuss the magnetic ($\mu$) and quadrupole moment $(Q)$ for $^{204-213}$Bi isotopes;  the corresponding comparisons with the experimental data are reported in Table \ref{qm}. In the calculation of quadrupole moments, we have used the effective charges  e$_p=$1.5e and e$_n=$0.5e. In the case of magnetic moments, we have taken the gyromagnetic ratios as $g_l^\pi$ = 1.00 and $g_l^\nu$ = 0.00 for orbital angular momenta, and $g_s^\pi$ = 3.910, $g_s^\nu$ = -2.678 for spin angular momenta.  
	With the standard values $g_s^{\pi}\approx5.585$ and $g_s^{\nu}\approx-3.826$, somehow, we found notably smaller magnetic moments compared to the experimental data. Consequently, we have taken quenched values $g_s^{eff} =0.7 \times g_s^{free}$ (3.910, -2.678).  Also generally used in previous studies for different mass regions (for e.g., in Refs. \cite{Stuchbery, Teruya1, deepak}). Utilizing these quenched values, we found that the magnitude of magnetic moments increases and approaches towards the experimental values, and the overall trends of $\mu$-moments also remain unchanged. It hints that utilizing quenched values is more relevant in the present work.
We have reported these properties using KHH7B interaction only.

For $^{204}$Bi, the calculated magnetic moments for $6^+_1$ and $10^-_1$ states show reasonable agreement with the corresponding experimental data. The quadrupole moment for the $6^+_1$ state is reproduced very well, i.e., -0.48 $eb$ (corresponding experimental value is -0.49(15) $eb$). However, for the $10^-$ state, we are getting the opposite sign in the calculated quadrupole moment (-0.05 $eb$). By considering the single-particle estimation of quadrupole moment for $10^-_1$ $[\pi(h_{9/2})\otimes\nu(i_{13/2}^{-1})]$ state and taking the assumption $Q(particle)=-Q(hole)$; we found that the $Q_{total}$ $(\approx e_{eff}^pQ(\pi h_{9/2})+e_{eff}^nQ(\nu i_{13/2}^{-1})$) has a negative sign corresponding to effective charges $(1.5,0.5)$. In the case of odd-$A$ $^{205-213}$Bi, the magnetic and quadrupole moment for the $9/2^-_1$ state is reproduced nearly their experimental values. The magnetic moment for  $25/2^+_1$ state is also reproduced well in $^{205}$Bi. In the case of $^{206}$Bi, the calculated magnetic moments for $6_1^+$  state shows reasonable agreement with the experimental data. We are able to reproduce the quadrupole moment for the $6^+_1$ state, which is -0.40 $eb$ (corresponding experimental value is -0.39(4) $eb$). Like $^{204}$Bi, quadrupole moment for $10^-_1$ $[\pi(h_{9/2})\otimes\nu(i_{13/2}^{-1})]$ state in $^{206}$Bi also shows a negative sign, whereas its experimental value has a positive sign. However, the calculated $Q(10^-_1)$ using the single-particle estimate with the effective charges $(1.5e,0.5e)$ supports the negative sign of the corresponding SM obtained result. In $^{207}$Bi, the magnetic moment for the $21/2^+$ state is underestimated by a factor of $\approx$1.5. In the case of $^{208}$Bi,  the calculated magnetic moment for $5_1^+$ is in reasonable agreement with its experimental value, whereas the magnetic moment of the $10^-_1$ state is underestimated by a factor $\approx$ 1.6. We are able to reproduce the quadrupole moment well for the $5^+_1$ state.  The calculated quadrupole moment for the $9/2^+_1$ state in $^{209}$Bi matches well with its experimental value. We are unable to reproduce the sign of experimental $Q$-moment for $15/2^+$ $[\pi(h_{9/2})\otimes\nu(p_{1/2}^{-1}g_{9/2}^1)]$ state in $^{209}$Bi. For $^{210}$Bi, the calculated magnetic moment for $9^-_1$, $7^-_1$, and $5^-_1$ is reproduced reasonably well. It is difficult to explain how to get the experimental sign of the magnetic moment of $1^-$ [$\pi(h_{9/2})\otimes\nu(g_{9/2})$] state in $^{210}$Bi. However, the empirical formula of $g$ factor ($g=(g(\pi h_{9/2})+g(\nu g_{9/2}))/2$) for the present configuration also suggests a positive sign of $\mu(1^-)$ similar to our case \cite{Baba}. The quadrupole moment for the $1^-_1$ and $9^-_1$ states are also reproduced very well. For $^{211}$Bi, experimentally, the magnetic moment for $9/2^-_1$ state has an unconfirmed positive sign; on comparing it with our calculated result, we can predict the sign to be positive. We are also able to reproduce the magnetic moment for the $7/2^-_1$ state near its experimental value. Our calculations reproduce magnetic and quadrupole moment for $1^{(-)}_1$ state quite well for $^{212}$Bi isotope. In Fig. \ref{quad_magnetic}, we can see that the observed magnetic moment increases for the 9/2$^-_{g.s.}$ state up to $N=126$, and it keeps decreasing for $N>126$, which also reflects in our calculations. However, our results are slightly underpredicted. The trend of $Q(9/2^-_{g.s.})$ in odd-$A$ Bi isotopes is also consistent with the experimental data as shown in Fig. \ref{quad_magnetic}.

According to the collective model, we can correlate the spectroscopic quadrupole moment $(Q_s)$ with the intrinsic quadrupole moment $(Q_0)$ by the following expression

\begin{equation} \label{Qs}
	Q_s=\frac{3K^2-I(I+1)}{(I+1)(2I+3)}Q_0,
\end{equation}

where $K$ is the projection of nuclear spin ($I$) along the symmetry axis.
The static deformation parameter $\beta_2\sim-0.03$ is reported in 
Ref. \cite{Kilgallon} for $^{213}$Bi.
To obtain the $K$ number, first we have to calculate $(Q_0)$ for $^{213}$Bi, using expression given below 

\begin{equation} \label{Q0}
	Q_0=\frac{3ZR^2}{\sqrt{5\pi}}\beta_2.
\end{equation}

Here, $R$ is taken as $1.2\times A^{1/3}$ fm. We have used the experimental value of $Q_s$ from two Refs. \cite{Kilgallon} and \cite{Bieron} to calculate $K$ number corresponding to nuclear spin $I^{\pi}=9/2^-_{g.s}$; the obtained $K$ are 4.68 and 5.21, respectively. For $\beta_2\approx -0.03$, the calculated $K$ numbers in $^{205-213}$Bi lie in the same range. From these interpretations, we can assume the suitable $K$ number as 4.5 for the nuclei of interest. Again, we have calculated the $Q_0$ values for the $9/2^-_{g.s}$ by taking $K=9/2$ and using the shell model predicted $Q_s$ values (here again we use Eq. (\ref{Qs})). We found that the $Q_0$ values keep the negative sign throughout the $^{205-213}$Bi isotopes. Since it is suggested in the Bohr-Mottelson model \cite{Bohr}, the negative sign of the intrinsic quadrupole moment $(Q_0)$ corresponds to the oblate deformation. Thus, we may predict the ground states of the odd-$A$ $^{205-213}$Bi isotopes exhibit oblate deformation.

Overall, the present shell-model calculation reasonably reproduces the experimental data for magnetic and quadrupole moments. We have also reported the electromagnetic moments of various states for which experimental data are not available. Our predictions could be useful for comparing data from upcoming experiments.

\subsection{\label{isomer1}Isomeric states}

\begin{table}
	\begin{center}
		\caption{Dominant Configurations (with KHH7B) of the isomers in the Bi isotopes with their seniority.}
		\label{t_sen}
		
		\begin{tabular}{rrcccc}
			\hline
			&  &  &   & &   \\
			Nucleus & $J^{\pi}$ & Seniority  & Wave-function & Probability  \\
			
			&  &  &   & &   \\
			\hline
			$^{204}$Bi &10$^-$&$v$=2 &$\pi$(h$_{9/2})$$\otimes$ $\nu$(i$_{13/2}^{-1}$)  & 49.12\\
			$ $ & 17$^+$&$v$=4 &$\pi$(h$_{9/2})$$\otimes$$\nu$(f$_{5/2}^{-1}$i$_{13/2}^{-2}$)  &63.89\\
			\hline
			$^{205}$Bi & 1/2$^+$& $v$=3 &$\pi$(h$_{9/2})$$\otimes$$\nu$(f$_{5/2}^{-1}$i$_{13/2}^{-1}$) &68.98\\
			$ $ & 21/2$^+$&$v$=3 & $\pi$(h$_{9/2})$$\otimes$$\nu$(f$_{5/2}^{-1}$i$_{13/2}^{-1}$)   & 59.43\\ 
			& 25/2$^+$& $v$=3&$\pi$(h$_{9/2})$$\otimes$$\nu$(f$_{5/2}^{-1}$i$_{13/2}^{-1}$)   &67.14\\
			\hline
			$^{206}$Bi & 4$^+$&$v$=2 & $\pi$(h$_{9/2})$$\otimes$$\nu$(f$_{5/2}^{-1}$)  & 60.48\\
			$ $ & 10$^-$&$v$=2 &$\pi$(h$_{9/2})$$\otimes$$\nu$(i$_{13/2}^{-1}$)  & 58.43\\
			$ $ & 15$^+$&$v$=4 &$\pi$(h$_{9/2})$$\otimes$$\nu$(p$_{1/2}$i$_{13/2}^{-2}$)   &82.51\\    
			\hline
			$^{207}$Bi &21/2$^+$&$v$=3  & $\pi$(h$_{9/2})$$\otimes$$\nu$(p$_{1/2}$i$_{13/2}^{-1}$)  & 90.18\\
			$ $ & 29/2$^-$& $v$=3 & $\pi$(h$_{9/2})$$\otimes$$\nu$(i$_{13/2}^{-2}$)   & 98.21\\
			\hline
			$^{208}$Bi & 10$^-$&$v$=2  &$\pi$(h$_{9/2})$$\otimes$$\nu$(i$_{13/2}^{-1}$)     & 99.48\\              
			\hline  
			$^{209}$Bi & 1/2$^+$&$v$=3 & $\pi$(h$_{9/2})$$\otimes$$\nu$(p$_{1/2}$g$_{9/2}$)   &92.86\\
			&  19/2$^+$&$v$=3 &$\pi$(h$_{9/2})$$\otimes$$\nu$(p$_{1/2}$g$_{9/2}$)   & 96.15\\
			\hline          
			$^{210}$Bi & {9}$^-$&$v$=2 & $\pi$(h$_{9/2})$$\otimes$$\nu$(g$_{9/2}$)   & 99.11\\
			&  $ $ {7}$^-$&$v$=2 & $\pi$(h$_{9/2})$$\otimes$$\nu$(g$_{9/2}$)   & 99.19\\
			& {5}$^-$&$v$=2 & $\pi$(h$_{9/2})$$\otimes$$\nu$(g$_{9/2}$)   & 99.27\\
			
			\hline
			$^{211}$Bi & {21/2}$^-$ &$v$=3 & $\pi$(h$_{9/2})$$\otimes$$\nu$(g$_{9/2}^2$)   & 98.04\\ 
			$ $ & {25/2}$^-$ &$v$=3 & $\pi$(h$_{9/2})$$\otimes$$\nu$(g$_{9/2}^2$) &97.75\\
			\hline    
			$^{212}$Bi & {8}$^-$ & $v$=2 &    $\pi$(h$_{9/2})$$\otimes$$\nu$(g$_{9/2}$) & 75.85\\
			$ $ & {9}$^-$ & $v$=2 &    $\pi$(h$_{9/2})$$\otimes$$\nu$(g$_{9/2}$) & 55.56\\
			& {18}$^-$&$v$=4 &   $\pi$(h$_{9/2})$$\otimes$$\nu$(i$_{11/2}$g$_{9/2}^2$)  & 98.84\\

			\hline 
			
		\end{tabular}
	\end{center}
\end{table}

There are several experimentally confirmed isomers in the Bi isotopes. Thus, it is highly desirable to interpret these isomeric states in terms of configuration and corresponding half-lives obtained from the shell model. The seniority, configurations, and half-lives of various isomeric states for Bi isotopes are described in Tables \ref{t_sen} and \ref{t_hl}. The internal conversion coefficients  \cite{Bricc} were used in the calculation of half-lives. The splitting of high-$j$ nucleon pairs in spherical nuclei around the magic number can result in isomeric states. The Bi isotopes that we have taken into consideration lie near the magic number $N=126$. Thus, it is possible to discuss isomers based on seniority quantum numbers. The seniority quantum number, denoted by $v$, is used to represent the number of unpaired nucleons,  which are not pair coupled to the angular momentum $J=0$. Using the shell-model, one can extract seniority information from configurations. The inhibition of decay in seniority isomers is a consequence of their initial and final states possessing the same seniority quantum number.

In $^{204}$Bi isotope $10^-_1$ [$\pi(h_{9/2}^1)$  $\otimes $  $\nu (f_{5/2}^4p_{3/2}^4i_{13/2}^{13})$] isomeric state decays to $7^+_1$ state by electric octupole transition $(E3)$, is coming due to coupling of one valance proton in $h_{9/2}$ and one neutron hole in $i_{13/2}$ orbital. Since this isomer has two unpaired nucleons, therefore seniority ($v$) of this state is 2. Another isomeric state in this isotope, $17^+_1$ [$\pi(h_{9/2}^1)$  $\otimes $  $\nu (f_{5/2}^5p_{3/2}^4i_{13/2}^{12})$] is formed due to coupling of one valance proton in $h_{9/2}$, one neutron hole in $f_{5/2}$ and two neutron hole in $i_{13/2}$ orbital, having seniority ($v$)= 4. This state decays to $15^-_1$, and $14^-_1$ via $M2$ and $E3$ transition, respectively.

In $^{206-208}$Bi isotope $10^-_1$ [$\pi(h_{9/2}^1)\otimes\nu (i_{13/2}^{-1})$] isomeric state is obtained from the coupling of one valance proton in $h_{9/2}$ and one neutron hole in the $i_{13/2}$ orbital, having seniority ($v$)=2. Experimentally in $^{206}$Bi, the $10^-_1$ isomer decays via $10^-_1\rightarrow 7^+_1$ $E3$ and $10^-_1\rightarrow 8^+_1$ $M2$ transitions with a half-life 0.89(1) ms \cite{NNDC}. The shell model predicted half-life (3.2 ms) is consistent with the experimental value with a small deviation. In $^{206}$Bi, the $4^+_1$ state decays via $E2$ transition into $6^+_{g.s}$; both states are coming from [$\pi(h_{9/2}^1)\otimes\nu (f_{5/2}^5p_{3/2}^4i_{13/2}^{14})$] configuration and favored by seniority $v=2$. Although, the calculated $B(E2;4^+_1\rightarrow6^+_1$) value is smaller (5.91$\times 10^{-6}$ W.u.) compared to the corresponding experimental value (0.0180(6) W.u.).  Our calculated half-life using $B(E2;4^+_1 \rightarrow 6^+_1)$ for $4^+_1$ isomer is large, i.e., 18.5 ms (corresponding experimental value is 7.7(2) $\mu s$). The 4$^+_1$ and 6$^+_1$ states both have the same seniority,  due to which the $4^+_1\rightarrow6^+_1$ transition is hindered and a small $B(E2;4^+_1\rightarrow 6^+_1)$ value is obtained. This supports that the $4^+$  isomer can be considered as a seniority isomer.
The $15^+_1$ is an isomeric state in both $^{206}$Bi and $^{210}$Bi. In $^{206}$Bi $15^+_1$ [$\pi(h_{9/2}^1)\otimes\nu(p_{1/2}^1i_{13/2}^{12})$] isomeric state stems from the one unpaired proton in $h_{9/2}$, one unpaired neutron in $p_{1/2}$ and two neutron holes in $i_{13/2}$ orbital, having seniority $v=4$. However, we are unable to calculate the $15^+_1$ state in the present calculation in the $^{210}$Bi isotope since we have not considered particle-hole excitation for this isotope.

In $^{210}$Bi isotope, $5^-_1$, $7^-_1$ and $9^-_1$ isomeric states having the same configuration $\pi(h_{9/2}^1)\otimes\nu (g_{9/2}^1)$ are formed due to coupling of one unpaired proton in $h_{9/2}$ and one unpaired neutron in $g_{9/2}$ orbital, thus seniority $v$ =2.  Experimentally, it was proposed that the $7^-_1$ isomer decays via $7^-_1\rightarrow9^-_1$ transition, while the relative intensities of $5^-_1\rightarrow3^-_1$ and $5^-_1\rightarrow7^-_1$ transitions were not observed \cite{Proetel}. Although the order of multipolarity is not known for both isomers yet, we have predicted the half-lives for the $5^-_1$ isomer using the $B(E2;5^-_1\rightarrow3^-_1)$ and $B(E2;5^-_1\rightarrow7^-_1)$ transitions and for $7^-_1$ isomer using $B(E2;7^-_1\rightarrow9^-_1)$ transition from KHH7B interaction. As reported in Table \ref{t_hl}, the shell model obtained half-lives for both isomers are not deviating much from the experimental values.

As reported in Ref. \cite{Chen}, the tentative $(8^-,9^-)$ isomeric state at 239(30) keV was observed in $^{212}$Bi. Analogous to this state, the $9^-$ isomer is also available in the same energy region \cite{Baisden} in $^{210}$Bi. In our shell-model calculations, both $9^-_1$ states (in $^{210,212}$Bi) coming due to the coupling of one unpaired proton in $h_{9/2}$ and one unpaired neutron in $g_{9/2}$ orbital having seniority $v=2$. By analyzing the above findings and the locations of the $8^-_1$ and $9^-_1$ states in $^{210,212}$Bi isotopes, we predict the shell-model calculated $9^-_1$ state could be a possible candidate for the experimental $(8^-,9^-)$ isomeric state in $^{212}$Bi. Another isomeric state $18^-_1$ [$\pi(h_{9/2}^1)\otimes\nu (i_{11/2}^1g_{9/2}^2)$] is obtained from the one unpaired proton in $h_{9/2}$, one unpaired neutron in $i_{11/2}$ and one neutron pair breaking in $g_{9/2}$ orbital  favored by seniority ($v$) equals to 4.

\begin{table*}
	\begin{center}
		\caption{The computed half-life  of isomeric states for Bi isotopes compared to the the experimental data (Expt.) \cite{NNDC,204BI,205BI,206BI,207BI,208BI,209BI,210BI,211BI,212BI,213BI}.}
		\label{t_hl}
		\begin{tabular}{cccccccc}
			\hline
			&   &  &   & & &  \\
			Isotope  &  $J^{\pi}_i$ & {$J^{\pi}_f$} & $E_{\gamma}$ & $B(E\lambda)$  & $B(E\lambda)$($e^2$fm$^{2\lambda}$)& Expt. & SM \\
			& & & (MeV)  & or $B(M\lambda)$  &$B(M\lambda)$($\mu_N^2$fm$^{2\lambda -2}$)   &T$_{1/2}$ & T$_{1/2}$  \\
			& & &  &   & &  &  \\
			\hline
			\hline
			
			& & &  &   & & &   \\
			$^{204}$Bi  & $10_1^-$ & {$7^+_1$} & 0.891 & $B(E3)$ & 1.2$\times 10^{-4}$  & 13.0(1) ms & 22.4 s  \\
			& $17_1^+$ & {$15^-_1$} & 0.511 & $B(M2)$  & 3.8$\times 10^{-2}$   & 1.07(3) ms & 29.3 $\mu$s  \\
			&  & {$14^-_1$} & 1.033 & $B(E3)$ & 2.6$\times 10^{-2}$  & &   \\
			
			& & &  &   & & &   \\
			
			\hline
			
			& &  &  &   & & &  \\
			$^{205}$Bi  &  $25/2_1^+$ & {$21/2^+_1$} & 0.053 & $B(E2)$ & 51.1   & 220(25) ns & 200.7 ns  \\
			&  $21/2_1^+$ & {$17/2^-_1$} & 0.802 & $B(M2)$ & 7.9$\times10^{-2}$   & 100(6) ns &  114.6 ns  \\
			&  & {$17/2^+_1$} & 0.052 & $B(E2)$ & 84.2   &  &   \\
			
			& & &  &   & & &   \\
			
			\hline
			
			& &  &  &   & & &   \\
			$^{206}$Bi  &  $4^+_1$ & {$6^+_1$} & 0.093 & $B(E2)$  &  $4.3\times10^{-4}$   & 7.7(2) $\mu$s & 18.5 ms   \\
			&  $10^-_1$ & {$7^+_1$} & 1.071 & $B(E3)$ & $0.1\times10^{-3}$ &  0.89(1) ms  & 3.2 ms  \\
			&  & {$8^+_1$} & 0.339 & $B(M2)$ & 1.7$\times 10^{-3}$ &  &  \\
			
			& & &  &   & & &   \\
			
			\hline
			
			& & &  &   & & &  \\
			$^{207}$Bi  &  $21/2_1^+$ & {$15/2^-_1$} & 0.880 & $B(E3)$ & 7.8$\times 10^{-5}$   & 182(6) $\mu$s & 33.1 s  \\
			&  & {$15/2^-_2$} & 0.426 & $B(E3)$ & 1.5$\times10^{-3}$    &               &         \\
			
			& & &  &   & & &   \\
			
			\hline
			
			& & &  &   & & &  \\
			
			$^{209}$Bi  &  $1/2_1^+$ & {$7/2^-_1$} & 2.421 & $B(E3)$ & 5.5  & 11.3(4) ns & 453.7 ns  \\
			&  $19/2_1^+$ & {$15/2^+_1$} & 0.220 & $B(E2)$ & 17.6   & 17.9(5) ns & 47.1 ns  \\
			
			& & &  &   & & &   \\
			
			\hline
			
			& & &  &   & & &   \\
			$^{210}$Bi  &  $7^-_1$ & {$9^-_1$} & 0.190 & $B(E2)$ & 19.3 & 57.5(10) ns & 77.0 ns  \\
			&  $5^-_1$ & {$3^-_1$} & 0.082 & $B(E2)$ & 127.7  & 38(1) ns & 61.1 ns  \\
			&  & {$7^-_1$} & 0.006 & $B(E2)$ & 96.2 & & \\
			
			& & &  &   & & &   \\
			
			\hline
			
			& & &  &   & & &   \\
			$^{211}$Bi  &  $21/2_1^-$ & {$17/2^-_1$} & 0.054 & $B(E2)$ & 53.2   & 70(5) ns & 192.2 ns \\
			&  $25/2_1^-$ & {$21/2^-_1$} & 0.031 & $B(E2)$ & 26.7   & 1.4(3) $\mu$s & 0.4 $\mu$s \\
			
			& & &  &   & & &   \\
			
			\hline
			
		\end{tabular}
	\end{center}
\end{table*}

In $^{205}$Bi, experimentally, the isomeric state $25/2^+_1$ decays into $21/2^+_1$ state via $E2$ transition with a half-life 220(25) ns \cite{NNDC}. The KHH7B interaction predicts that both states are originating from the same dominant configuration [$\pi(h_{9/2}^1)\otimes \nu(f_{5/2}^{-1}i_{13/2}^{-1})_{6^-,8^-}$] and favored by seniority $v=3$. Despite having the same configuration, our calculations give a small $B(E2)$ value (0.7 W.u.) similar to the experimental value (see Table \ref{be2}); the obtained half-life ($T_{1/2}=200.7$ ns) is also quite close to the experimental half-life. Here, the $B(E2;25/2^+\rightarrow 21/2^+)$ transition is hindered due to the same seniority. On the basis of the above discussion, we can categorize the $25/2^+$ state as a seniority isomer. Experimentally, the isomeric state $21/2^+_1$ decays to $17/2^+_1$ and $17/2^-_1$ state via $E2$ and $M2$ transitions, respectively \cite{NNDC}. Although the $B(E2;21/2^+_1\rightarrow 17/2^+_1)$ value is not observed yet, our calculation using KHH7B interaction shows the dominant branching fraction (93.7\%) from this transition. Our calculated total half-life of the $21/2^+$ isomer [$\pi(h_{9/2}^1)\otimes\nu(f_{5/2}^{-1}i_{13/2}^{-1})$] using $B(E2)$ and $B(M2)$ values shows quite good agreement with the experimental data (see Table \ref{t_hl}).

In $^{207}$Bi isotope, the ${21/2}^+_1$ [$\pi(h_{9/2}^1)\otimes\nu (p_{1/2}^{-1}i_{13/2}^{-1})$] isomeric state decays to $15/2_1^-$ and  $15/2_2^-$ states via $E3$ transition, is formed due to coupling of one unpaired proton in $h_{9/2}$ orbital, one neutron hole in $p_{1/2}$, and one neutron hole in $i_{13/2}$ orbital, having seniority ($v$)=3. Our SM obtained $B(E3;{21/2}^+_1\rightarrow15/2^-_1)$, and $B(E3;{21/2}^+_1\rightarrow15/2^-_2)$ values are very small i.e. 3.06$\times10^{-8}$, and 5.89$\times10^{-7}$ W.u. (corresponding experimental values are 0.33(5), and 0.008311 W.u.), due to which the computed half-life of this state is very large. Another isomeric state in this isotope is $29/2^-_1$ [$\pi(h_{9/2}^1)\otimes\nu(i_{13/2}^{-2})$] formed due to coupling of one unpaired proton in $h_{9/2}$, and two neutron hole in $i_{13/2}$ orbital, as a consequence seniority ($v$)=3.

In $^{209}$Bi isotope $1/2^+_1$ and $19/2^+_1$ both isomeric states have configuration $\pi(h_{9/2}^1)\otimes\nu(p_{1/2}^1 g_{9/2}^1)$. Thus, these isomeric states are obtained from the coupling of one unpaired proton in $h_{9/2}$, one unpaired neutron in each $p_{1/2}$ and $g_{9/2}$ orbital, hence both of these states exhibit seniority ($v$) 3. The isomeric state $1/2^+_1$ decays to $7/2^-_1$ state by $E3$ transition, therefore we have calculated its half-life using $B(E3)$ value, which comes to be 455.0 ns (corresponding experimental half-life is 11.3(4) ns). We are able to reproduce the half-life for isomeric state $19/2^+_1$ to the same order as the experimental value, i.e., 46.8 ns (corresponding experimental value is 17.9(5) ns) using $B(E2;19/2^+_1 \rightarrow 15/2^+_1)$ value. Both states are characterized by the same dominant configuration $\pi(h_{9/2}^1)\otimes \nu(p_{1/2}^1 g_{9/2}^1)$. We obtain a small value of $B(E2;19/2^+_1 \rightarrow 15/2^+_1)$ due to the same seniority ($v$), i.e., 3 for both states. The above results indicate that the $19/2^+_1$ state can be classified as a seniority isomer.

Experimentally in $^{211}$Bi, the $21/2^-$ and $25/2^-$ isomers decay via $21/2^-_1\rightarrow 17/2^-_1$ and $25/2^-_1\rightarrow 21/2^-_1$ $E2$ transition \cite{NNDC}. The related states ($17/2^-_1$, $21/2^-_1$, and $25/2^-_1$) are characterized by same dominant configuration $\pi(h_{9/2}^1)\otimes \nu(g_{9/2}^2)$ and having seniority $v=3$. The KHH7B interaction predicts weak $E2$ transition between these states; the calculated half-lives of $21/2^-_1$ and $25/2^-_1$ states using the shell-model obtained $B(E2)$ transitions are showing reasonable agreement with the experimental values (see Table \ref{t_hl}). Our results favor that both isomeric states might be seniority isomers.

\subsection{The root mean square deviation} \label{rms_deviation}
We have calculated the root mean square (rms) deviation between the SM obtained result and the experimental data, by utilizing the following expression

\begin{equation}
	rms=\sqrt{\frac{1}{N}\sum_{i=1}^N(E_{exp}^i-E_{th}^i)^2}. 
\end{equation}

Here, $E^i_{exp}$ and $E^i_{th}$ represent the experimental and theoretical observables, respectively. The rms deviation  in SM obtained values (with KHH7B interaction) such as reduced transition probability $B(E2)$, quadrupole, and magnetic moment have been calculated using the data shown in Table \ref{be2} and \ref{qm}.  We have omitted those states for the rms deviation calculation for which experimental data are unavailable. The rms deviation is 2.332$^{+1.315}_{-1.099}$ W.u. for $B(E2)$, and $1.201^{+0.280}_{-0.248}$ $\mu_N$ for the magnetic moment. The rms deviation in quadrupole moments ($Q$) without considering experimental uncertainties is determined as 0.141 $eb$. By adding and subtracting maximum uncertainties in the experimental $Q$-moments, the obtained rms deviations are larger than 0.141 $eb$ by 0.076 and 0.073 $eb$, respectively. We obtain a relatively small rms deviation for quadrupole moments. It shows better predictive efficiency of the employed effective interaction and corresponding wave function for the $Q$-moments. Although, the rms deviation for magnetic moments ($\mu$) is slightly larger. Looking at the trend of $\mu$-moments, we can say that the slightly lower $\mu$-moments by shell model could be due to the choice of the quenched $g_{s}^{\pi}$ and $g^{\nu}_s$ values.


\section{Summary and conclusion\label{IV}}


In the present study, we have investigated the $^{204-213}$Bi isotopes to study the nuclear structure properties such as energy spectra, transition probabilities, electromagnetic moments, and nuclear isomers in the framework of shell model using KHH7B and KHM3Y effective interactions. Overall, the present shell-model results for the energy spectra in $^{204-213}$Bi isotopes exhibit satisfactory consistency with the experimental data. From this study, we have observed that high-lying states exhibit less collective behavior, whereas there is a significant configuration mixing in the low-lying states. Additionally, we have reported shell-model results corresponding to the electromagnetic properties for selected states for which no experimental data is available. Thus, it will be quite helpful to compare them with future experimental results. Based on shell-model interpretation for the intrinsic quadrupole moments of $9/2^-_{g.s}$, we have also predicted that the ground states of the odd-$A$ Bi isotopes show oblate deformation. We have interpreted various isomeric states and calculated their half-lives. Using the shell model configurations, reduced transition probabilities, and seniority of the initial and final states, we have categorized some of these isomers as seniority isomers.


\section*{{Acknowledgements}}
We are thankful for the financial assistance provided by MHRD, the Government of India, and SERB (India), CRG/2022/005167. We acknowledge the National Supercomputing Mission (NSM) for providing computing resources of ‘PARAM Ganga’ at the IIT Roorkee, implemented by C-DAC and supported by the Ministry of Electronics and Information Technology (MeitY) and Department of Science and Technology (DST), Government of India. We would also like to thank Professor Larry Zamick for useful discussions during this work.


\end{document}